\documentclass[11pt]{article}

\usepackage[final]{acl}

\usepackage{times}
\usepackage{latexsym}

\usepackage[T1]{fontenc}

\usepackage[utf8]{inputenc}

\usepackage{microtype}

\usepackage{inconsolata}

\usepackage{graphicx}

\title{MSVBench: Towards Human-Level Evaluation of Multi-Shot Video Generation}

\usepackage{authblk}

\author[1]{Haoyuan Shi\thanks{Email: 24s051044@stu.hit.edu.cn}}
\author[1]{Yunxin Li}
\author[1]{Nanhao Deng}
\author[1]{Zhenran Xu}
\author[1]{ \\Xinyu Chen}
\author[2]{Longyue Wang}
\author[1]{Baotian Hu\thanks{Corresponding author.}}
\author[1]{Min Zhang}

\affil[1]{Harbin Institute of Technology (Shenzhen)}
\affil[2]{Alibaba International Digital Commerce}

\date{} 

\usepackage{makecell}
\usepackage{multirow}
\usepackage{booktabs}
\usepackage[utf8]{inputenc}
\usepackage{graphicx}
\usepackage[table,xcdraw]{xcolor}
\usepackage{geometry}
\usepackage{amsmath}

\begin{document}
\maketitle
\begin{abstract}
The evolution of video generation toward complex, multi-shot narratives has exposed a critical deficit in current evaluation methods. Existing benchmarks remain anchored to single-shot paradigms, lacking the comprehensive story assets and cross-shot metrics required to assess long-form coherence and appeal. To bridge this gap, we introduce \textbf{MSVBench}, the first comprehensive benchmark featuring hierarchical scripts and reference images tailored for \textbf{M}ulti-\textbf{S}hot \textbf{V}ideo generation. We propose a hybrid evaluation framework that synergizes the high-level semantic reasoning of Large Multimodal Models (LMMs) with the fine-grained perceptual rigor of domain-specific expert models. Evaluating 20 video generation methods across diverse paradigms, we find that current models—despite strong visual fidelity—primarily behave as visual interpolators rather than true world models. We further validate the reliability of our benchmark by demonstrating a state-of-the-art Spearman’s rank correlation of \textbf{94.4\%} with human judgments. Finally, MSVBench extends beyond evaluation by providing a scalable supervisory signal. Fine-tuning a lightweight model on its pipeline-refined reasoning traces yields human-aligned performance comparable to commercial models like Gemini-2.5-Flash.
\end{abstract}

\begin{figure*}[t]
    \centering
    \includegraphics[width=1\textwidth]{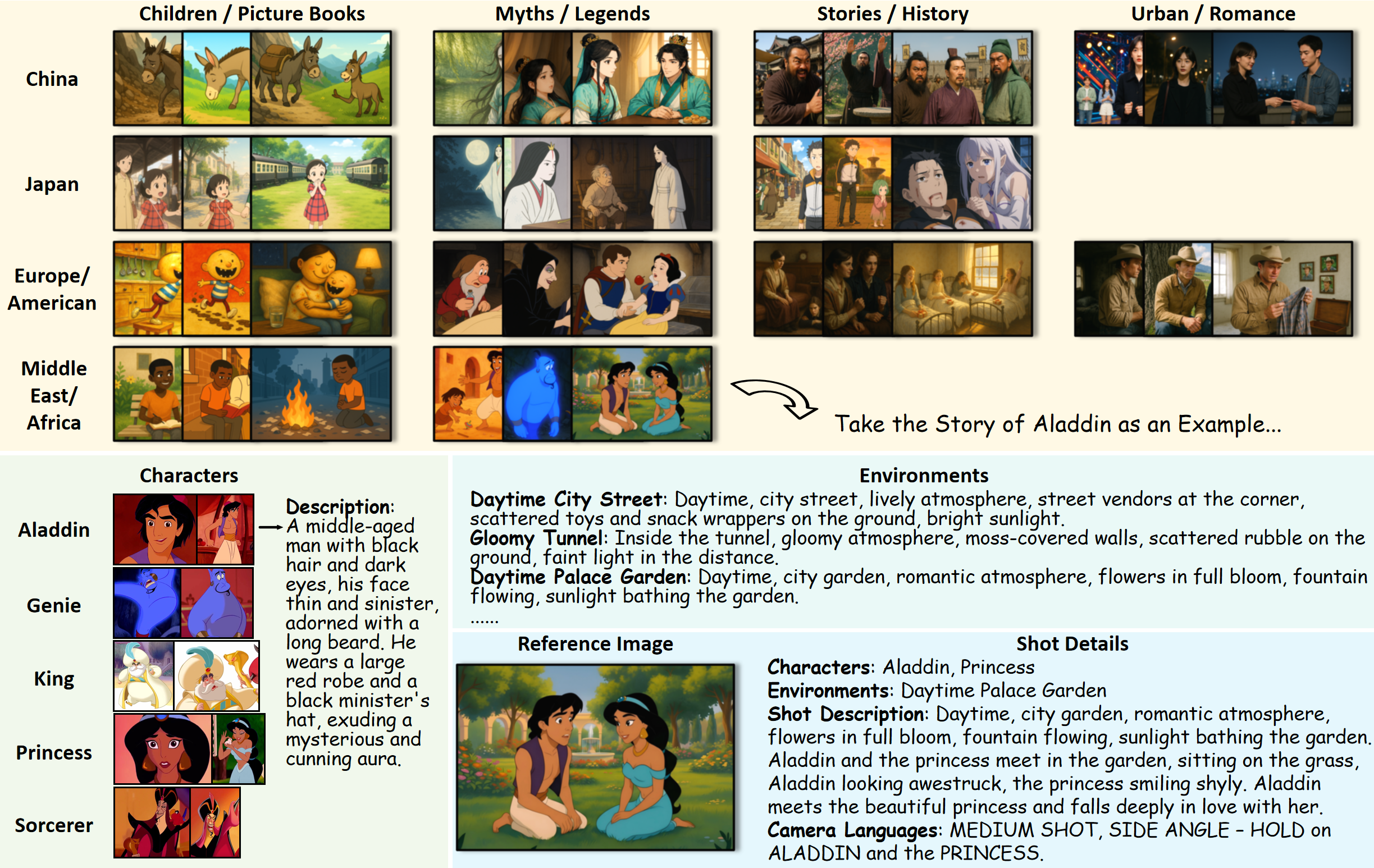} 
    \caption{The Hierarchical Data Organization of MSVBench.}
    \label{fig:Dataset}
\end{figure*}

\section{Introduction}

The field of video generation is transitioning from isolated, short clips to complex, multi-shot narratives, poised to revolutionize industries ranging from film production to interactive gaming. Yet, while commercial pioneers like Sora and Veo 3 propel the community toward world models, evaluation methods lag significantly behind these generative breakthroughs, creating a critical bottleneck.

Existing benchmarks exhibit fundamental limitations when applied to complex multi-shot video evaluation. Early efforts such as VBench \cite{huang2024vbench} and EvalCrafter \cite{liu2024evalcrafter} depend on lightweight expert models with limited video-understanding ability, leading to unreliable assessments for intricate actions or narrative-rich content. More recent attempts, including Video-Bench \cite{han2025video}, adopt LMMs to approximate human reasoning; however, their exclusive reliance on LMMs introduces a lack of objective, standardized criteria and insufficient domain-specific perceptual grounding. More importantly, these benchmarks remain anchored to single-shot prompt–video pairs and are therefore misaligned with the emerging paradigm of multi-shot, story-driven video generation.

Although pioneers like OpenS2V-Nexus \cite{yuan2025opens2v} and ViStoryBench \cite{zhuang2025vistorybench} take initial steps toward story-level video evaluation, they still face two structural limitations that are particularly critical for multi-shot video evaluation. First, their story assets remain incomplete: the absence of fully detailed scripts and per-shot reference images limits the diversity of generation paradigms that can be evaluated. Second, their metric suites do not yet capture essential shot-level and cross-shot properties—such as temporal logic across consecutive shots—resulting in insufficient coverage of the core challenges unique to multi-shot narratives. Consequently, despite meaningful progress, these benchmarks cannot yet replace the costly and non-scalable practice of human evaluation in multi-shot video generation.

To address this, we introduce \textbf{MSVBench}, a comprehensive framework specially designed for multi-shot video evaluation. We formulate the task hierarchically, decomposing stories into global priors, scene-level segments, and shot-level conditions. To achieve human-level precision, our framework synergizes the semantic reasoning of LMMs with the perceptual fidelity of domain-specific expert models. This hybrid design enables a unified evaluation that jointly captures low-level visual fidelity and high-level narrative coherence, achieving a record-high Spearman’s rank correlation of 94.4\% with human judgments.

We conduct an in-depth evaluation of 20 diverse generation methods, ranging from commercial leaders to open-source and agent-based paradigms. Results reveal that while top-tier commercial models excel in dimensions like motion quality, open-source contenders are rapidly narrowing the gap. Furthermore, we expose a fundamental limitation in current video generation frameworks: Despite high prompt alignment capacity, current systems struggle with long-form stability and consistency across shots, functioning more as visual interpolators than true world models.

Finally, we construct a data pipeline that converts evaluation traces into thinking process data for high-quality supervision. By fine-tuning a lightweight Qwen3-VL-4B model on this data, we achieve human-aligned performance that surpasses commercial models like Gemini-2.5-Flash.

Our contributions are threefold:
\begin{itemize}
    \item We propose \textbf{MSVBench}, the first comprehensive benchmark tailored for multi-shot video generation. MSVBench features a hierarchical data formulation and a hybrid evaluation framework, harmonizing the semantic reasoning capabilities of LMMs with the perceptual precision of domain-specific expert models.
    \item We evaluate 20 diverse systems, validating our benchmark's reliability with a state-of-the-art 94.4\% human correlation. The results expose that current models operate primarily as visual interpolators rather than true world models.
    \item We construct a systematic pipeline transforming evaluation traces into supervision data. Training a specialized lightweight model on this data yields human-aligned performance surpassing even commercial models.
\end{itemize}

\section{Related Work}

\begin{table*}[h]
\centering
\footnotesize 
\caption{Comparison of video benchmarks. \textbf{Ref(C/S)}: Character/Shot reference images. \textbf{Aspects}: Visual Quality (VQ), Story/Prompt Video Alignment (SVA/PVA), Video Consistency (VC), Motion Quality (MQ). \textbf{Evaluator}: Evaluation backbone (Spec.: Specialized Models). \textbf{H-Align}: Correlation analysis with human evaluation.}

\label{tab:video_benchmarks}
\setlength{\tabcolsep}{3pt} 
\begin{tabular}{lccccccc}
\toprule
\textbf{Benchmark} & \textbf{Ref(C/S)} & \textbf{Aspects} & \textbf{Evaluator} & \textbf{H-Align} & \textbf{Metrics} & \textbf{Methods} \\
\midrule
VBench\cite{huang2024vbench} & -/- & VQ/PVA/VC/MQ & Spec. & \textbf{Yes} & 16 & 9 \\
EvalCrafter\cite{liu2024evalcrafter} & -/- & VQ/PVA/VC/MQ & Spec. & \textbf{Yes} & 16 & 8 \\
Video-Bench\cite{han2025video} & -/- & VQ/PVA/VC/MQ & LLM & \textbf{Yes} & 9 & 7 \\
OpenS2V-Nexus\cite{yuan2025opens2v} & 479/- & VQ/PVA/VC/MQ & LMM/Spec. & No & 19 & 8 \\
ViStoryBench\cite{zhuang2025vistorybench} & \textbf{509}/- & VQ/SVA/VC & LMM/Spec. & Partial & 12 & 18 \\
\midrule
\textbf{MSVBench} & 136/\textbf{276} & \textbf{VQ/SVA/VC/MQ} & \textbf{LMM+Spec.} & \textbf{Yes} & \textbf{20} & \textbf{20} \\ 
\bottomrule
\end{tabular}
\end{table*}

\subsection{Multi-shot Video Generation}
\label{sec:multiscene_gen}
Existing approaches fall into four streams: (1) \textbf{Storyboard-Driven Synthesis} decouples narrative from dynamics via a two-stage pipeline. Methods like StoryDiffusion \cite{zhou2024storydiffusion} and StoryAdapter \cite{mao2024story} first generate coherent keyframes, subsequently animated by advanced image-to-video models \cite{wan2025wan, yang2024cogvideox, kong2024hunyuanvideo, hacohen2024ltx, jiang2024anisora, huang2025self}. (2) \textbf{Agent-Based Frameworks} leverage LLM orchestration for planning. MovieAgent \cite{wu2025automated}, MM-StoryAgent \cite{xu2025mm}, and VideoGen-of-Thought \cite{zheng2024videogen} focus on structural reasoning, while AnimDirector \cite{li2024anim} and AniMaker \cite{shi2025animaker} employ a "generate-and-select" mechanism for quality assurance. Recent works like FilmAgent \cite{xu2025filmagent}, FilMaster \cite{huang2025filmaster}, and HoloCine \cite{meng2025holocine} further incorporate 3D environments or cinematic principles. (3) \textbf{Continuous Generation Techniques} maintain long-term coherence via autoregressive mechanisms (e.g., LongLive \cite{yang2025longlive}, RollingForcing \cite{liu2025rolling}) or test-time optimization (TTT-Video \cite{dalal2025one}). (4) \textbf{Commercial Solutions} such as Sora \cite{sora2024} and Veo 3 \cite{google_veo3_2025} set industry standards, demonstrating exceptional consistency across complex narratives.

\subsection{Video Evaluation}
The landscape of video evaluation has evolved rapidly to keep pace with generative advancements. Prior reference-free benchmarks, such as VBench \cite{huang2024vbench} and EvalCrafter \cite{liu2024evalcrafter}, utilized specialized small models to assess basic quality dimensions. Recognizing the need for semantic reasoning, subsequent works like Video-Bench \cite{han2025video} shifted towards LMMs to better align with human perception. More recently, the focus has expanded to story-oriented generation. Benchmarks like OpenS2V-Nexus \cite{yuan2025opens2v} introduced reference image integration, while ViStoryBench \cite{zhuang2025vistorybench} pioneered story-level assessment. Table \ref{tab:video_benchmarks} presents a detailed comparison between these baselines and our MSVBench.

\section{MSVBench}

\subsection{Hierarchical Dataset Schema}
To evaluate multi-shot videos, MSVBench organizes data into a structured hierarchy comprising global priors, scene-level segments, and shot-level conditions. This schema accommodates diverse generation paradigms (Sec.~\ref{sec:multiscene_gen}) through the following components:

\noindent\textbf{Global Context.} We define global assets as a set of $n$ characters $\mathcal{C}=\{C_1, \dots, C_n\}$ and $k$ environments $\mathcal{E}=\{E_1, \dots, E_k\}$. Each character $C_i$ is represented by a tuple $(T_{name}, T_{desc}, \mathcal{I}_{ref})$, ensuring identity consistency via reference images. Similarly, each environment $E_j$ consists of a unique designation and a textual setting description.

\noindent\textbf{Hierarchical Script.} The narrative is structured as a sequence of \textit{scenes}, each anchored to a specific environment $E_i \in \mathcal{E}$. Each scene represents a continuous narrative segment, which is further decomposed into a sequence of atomic \textit{shots}.

\noindent\textbf{Shot Annotations.} Each shot $S_t$ contains comprehensive multimodal annotations: (i) \textit{Visual Context}, specifying the subset of on-screen characters $\mathcal{C}_{sub} \subseteq \mathcal{C}$ alongside a reference frame; (ii) \textit{Shot Description}, detailing visual states and dynamic actions; and (iii) \textit{Cinematography}, providing instructions for camera movements.

\subsection{Dataset Construction}
Derived from 20 diverse stories in ViStoryBench~\cite{zhuang2025vistorybench}, we restructure raw scripts into our hierarchical schema (see Fig.~\ref{fig:Dataset}). The data curation pipeline comprises three stages:

\noindent\textbf{Visual Grounding.} We synthesize high-fidelity reference frames via \texttt{GPT-Image-1} and \texttt{Nano Banana}, establishing consistent visual narratives.

\noindent\textbf{Prompt Refinement.} Text prompts are rewritten to jointly describe static states and dynamic actions, ensuring strict alignment with reference visuals.

\noindent\textbf{Cinematography Enrichment.} Leveraging \texttt{Gemini-2.5-Flash}, we translate static shot specifications (e.g., scale, angle) into explicit, dynamic camera motion instructions.

\begin{table}[h]
\centering
\caption{Overview of MSVBench Metrics.}
\label{tab:metric_summary}

\resizebox{\columnwidth}{!}{
\begin{tabular}{@{}p{1.8cm} p{3.8cm} p{6.4cm}@{}}
\toprule
\textbf{Dimension} & \textbf{Metric} & \textbf{Underlying Models} \\ \midrule
\multirow{4}{*}{\makecell[l]{Visual\\Quality}}
 & Dover Score & DOVER (VQA\_A, VQA\_T) \\
 & MusIQ Score & MusIQ \\
 & Visual Attr. Consist. & - \\
 & Style Consist. & CSD-ViT-L \\ \midrule
\multirow{5}{*}{\makecell[l]{Story\\Video\\Alignment}}
 & VQAScore & CLIP-FlanT5-XXL (VQAScore) \\
 & Detect \& Count Score & Gemini-2.5-Flash \\
 & Shot Perspective Align. & Gemini-2.5-Flash \\
 & State Shift \& Persistence & Gemini-2.5-Flash \\
 & Story Video Consist. & ShareCaptioner + KaLM-Embedding-V2 \\ \midrule
\multirow{5}{*}{\makecell[l]{Video\\Consistency}}
 & Face Consist. & \makecell[l]{SAM-Track + DeepFace / InceptionNeXt \\ + Gemini-2.5-Flash} \\
 & Character Consist. & \makecell[l]{SAM-Track + InceptionNeXt \\ + Gemini-2.5-Flash} \\
 & Background Consist. & Step1X-Edit + DreamSim \\
 & Clothes \& Color Consist. & Gemini-2.5-Flash \\
 & Relative Size Consist. & Gemini-2.5-Flash \\ \midrule
\multirow{5}{*}{\makecell[l]{Motion\\Quality}}
 & Action Recognition & \makecell[l]{SAM2 Tracker + VideoMAE V2 + CLIP \\ + Gemini-2.5-Flash} \\
 & Action Strength & RAFT \\
 & Camera Control & MonST3R + Gemini-2.5-Flash \\
 & Phys. Plausibility & Gemini-2.5-Flash \\
 & Phys. Interaction & Gemini-2.5-Flash \\

\bottomrule
\end{tabular}
}
\end{table}

\subsection{Metrics}

\begin{figure*}[t]
    \centering
    \includegraphics[width=1\textwidth]{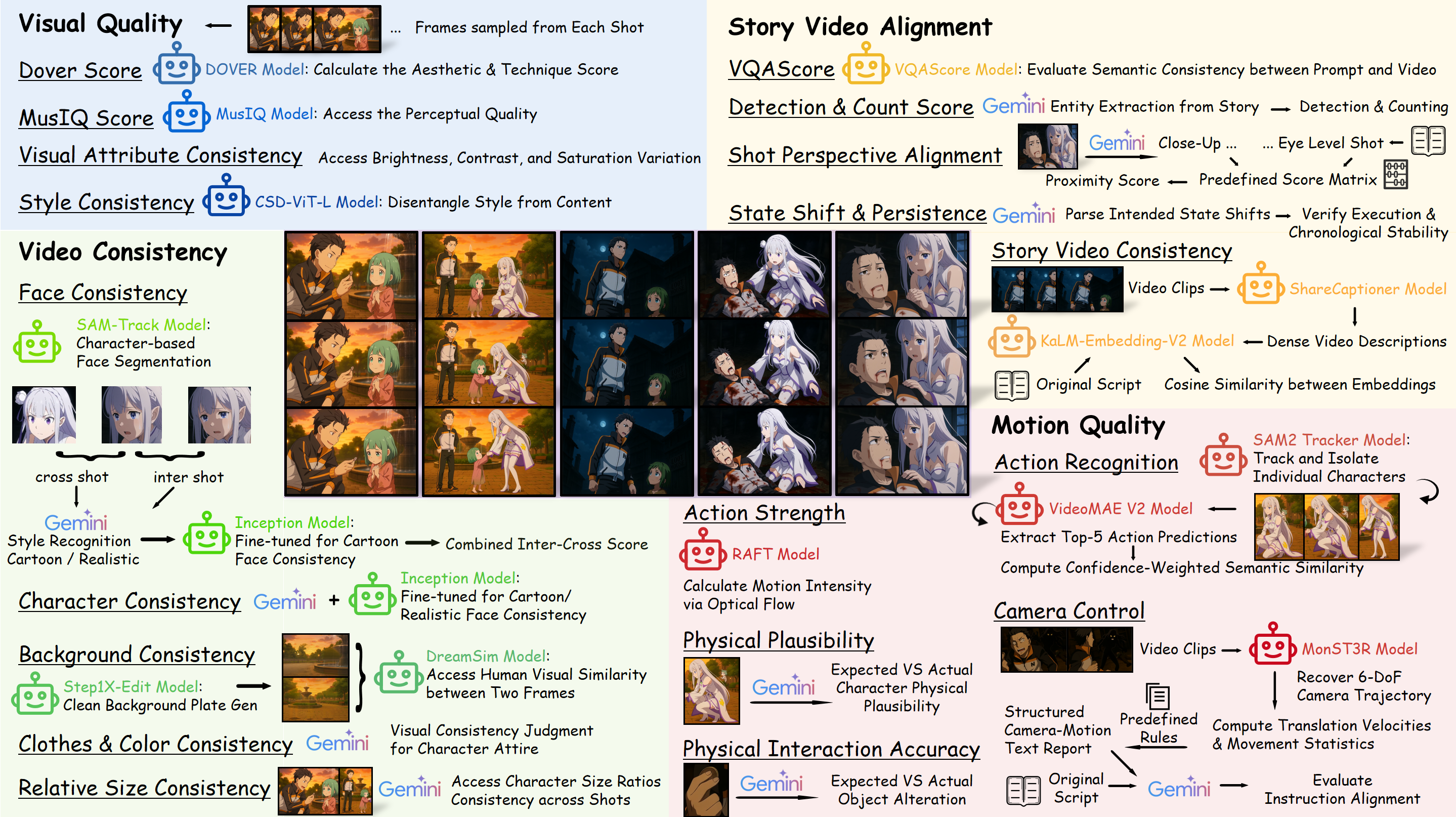} 
    \caption{The MSVBench Evaluation Framework.}
    \label{fig:Pipeline}
\end{figure*}

To bridge the gap between the limited semantic understanding of specialized small models and the lack of domain-specific precision in pure LMM approaches, we propose a hybrid evaluation framework, as shown in Figure \ref{fig:Pipeline}. This approach synergizes domain-specific expert models (e.g., DOVER, RAFT) for low-level perceptual fidelity with \texttt{Gemini-2.5-Flash} for high-level semantic reasoning. As summarized in Table \ref{tab:metric_summary}, we construct a comprehensive evaluation framework consisting of \textbf{20 sub-metrics} across \textbf{four dimensions}—Visual Quality, Story Video Alignment, Video Consistency, and Motion Quality—to systematically assess the capabilities of multi-shot long video generation. The detailed prompts for LMM are provided in Appendix~\ref{sec:MSVBenchPrompt}.

\subsubsection{Visual Quality}

\paragraph{Dover Score}
DOVER \cite{wu2023dover} is employed to assess video quality via VQA\_A (aesthetic appeal, e.g., composition) and VQA\_T (technical distortions, e.g., noise).

\paragraph{MusIQ Score}
MusIQ \cite{ke2021musiq} provides a unified perceptual index by averaging frame-level composition, sharpness, and artifact assessments.

\paragraph{Visual Attribute Consistency}
This metric evaluates the stability of brightness (mean intensity), contrast (intensity standard deviation), and saturation (mean HSV S-channel) by calculating their absolute differences across targeted frame pairs (e.g., adjacent frames within a clip or boundary frames between clips).

\paragraph{Style Consistency}
CSD-ViT-L \cite{somepalli2024measuring} is employed to extract style-disentangled embeddings at 1 fps, and the consistency score is defined as their mean cosine similarity across inter- and intra-shot frames.

\subsubsection{Story Video Alignment}

\paragraph{VQAScore}
To assess semantic consistency, we utilize VQAScore \cite{lin2024evaluating} with the CLIP-FlanT5-XXL backbone. Unlike standard embeddings matching, this metric employs a visual question-answering formulation to quantify text-frame alignment, averaged across frames sampled at 0.5s intervals.

\paragraph{Detection \& Count Score}
This metric verifies the generation of script-required objects and characters. Leveraging \texttt{Gemini-2.5-Flash}, we first extract target entities from the input prompt and then detect their actual presence in the video. The final score quantifies the inclusion of these mandated elements, accounting for their visibility duration.

\paragraph{Shot Perspective Alignment}
We leverage \texttt{Gemini-2.5-Flash} to align intended shot attributes with video realizations. Using a compatibility matrix over binary matching, the score reflects semantic proximity, penalizing severe deviations (e.g., Close-Up vs. Long Shot) more than minor shifts (e.g., Full Shot vs. Long Shot).

\paragraph{State Shift \& Persistence}
To assess chronological continuity, we employ a two-stage pipeline using \texttt{Gemini-2.5-Flash}. The model first identifies script-driven state changes (e.g., character appearance or environment) and localizes their occurrence in the video. Then it examines the \textbf{persistence} of the new state, aggregating scores for both the initial shift's execution and its subsequent stability.

\paragraph{Story Video Consistency}
To assess script-video consistency, we generate dense video descriptions via ShareCaptioner \cite{chen2024sharegpt4video}. These descriptions are compared against the input script within the KaLM-Embedding-V2 \cite{zhao2025kalm} semantic space, with the final score defined as the cosine similarity of their embeddings.

\subsubsection{Video Consistency}

\paragraph{Face Consistency} We quantify facial consistency using a style-adaptive pipeline. Following localization via SAM-Track \cite{cheng2023segment}, \texttt{Gemini-2.5-Flash} routes inputs to domain-specific extractors: DeepFace \cite{taigman2014deepface} for realistic content, and our custom InceptionNeXt \cite{yu2024inceptionnext} variant fine-tuned on a composite of anime datasets \cite{AnimeFaceDatasetKaggle,naftali2022aniwho,zheng2020cartoon} for animation. The final score is calculated as the feature distance between extracted embeddings.

\paragraph{Character Consistency} Adopting a similar pipeline, we evaluate character consistency by processing SAM-Track outputs via our specialized InceptionNeXt extractors. Guided by style identification from \texttt{Gemini-2.5-Flash}, the system selects between two variants we specifically fine-tune: a realistic model adapted from MVHumanNet \cite{xiong2024mvhumannet}, and an animated model trained on Hunyuan3D-1.0 \cite{yang2024hunyuan3d} synthetic data. The final score is derived from the embedding feature distance.

\paragraph{Background Consistency}
To assess environmental stability, we employ an occlusion-aware pipeline. Utilizing Step1X-Edit \cite{liu2025step1x} to remove foreground entities (e.g., characters) and yield clean background plates, we quantify their perceptual consistency via the DreamSim \cite{fu2023dreamsim} metric based on feature distance.

\paragraph{Clothes \& Color Consistency}
We evaluate attire stability using \texttt{Gemini-2.5-Flash} via a two-step protocol: confirming character presence before grading clothing consistency on a 5-point scale. Notably, the metric is robust to view changes, distinguishing true style discrepancies from natural variations caused by lighting or angles.

\paragraph{Relative Size Consistency}
Perspective-invariant size ratios between character pairs is estimated by \texttt{Gemini-2.5-Flash}. This metric isolates actual body dimensions from viewing angles, quantifying consistency via the temporal variance of these pairwise proportions.

\subsubsection{Motion Quality}

\paragraph{Action Recognition}
We adopt a character-centric approach for complex scenarios. Following isolation via \texttt{Gemini-2.5-Flash} and SAM2 Tracker \cite{ravi2024sam2}, VideoMAE V2 \cite{wang2023videomae} generates top-5 action predictions from the respective spatiotemporal regions. The consistency score is derived from the confidence-weighted sum of CLIP cosine similarities between these predictions and character-specific targets.

\paragraph{Action Strength}
To quantify motion intensity, we utilize RAFT \cite{teed2020raft} to extract dense optical flows between consecutive frames. The final score is defined as the average flow magnitude across the sequence, explicitly penalizing overly static content.

\paragraph{Camera Control}
To assess camera control fidelity, we employ MonST3R \cite{zhang2024monst3r} to recover frame-wise 6-DoF trajectories. These physical parameters are then translated into semantic motion descriptors (e.g., "pan right"), which \texttt{Gemini-2.5-Flash} compares against textual instructions to quantify execution accuracy.

\paragraph{Physical Plausibility}
We employ a character-centric pipeline powered by \texttt{Gemini-2.5-Flash} to verify adherence to Newtonian mechanics (e.g., gravity, momentum). The process involves extracting expected characters and dynamics from the prompt, followed by grading their motion realism in the video on a 5-point scale. The final metric averages these scores, explicitly penalizing violations like floating, interpenetration, or unnatural balance.

\paragraph{Physical Interaction Accuracy}
Complementing character-centric metrics, this module targets object-centric interaction fidelity. Utilizing \texttt{Gemini-2.5-Flash}, we extract expected reactions (e.g., deformation) from the prompt and verify their adherence to causal logic and material properties in the video. The score reflects the physical accuracy of consequential motions and structural changes.

\section{Experiments}

\definecolor{rankA}{RGB}{236, 112, 20}  
\definecolor{rankB}{RGB}{241, 132, 50}  
\definecolor{rankC}{RGB}{246, 152, 80}  
\definecolor{rankD}{RGB}{250, 172, 110} 
\definecolor{rankE}{RGB}{253, 192, 140} 
\definecolor{rankF}{RGB}{254, 212, 170} 
\definecolor{rankG}{RGB}{254, 232, 200} 
\definecolor{rankH}{RGB}{255, 245, 230} 
\newcommand{\rankbox}[1]{\textcolor{#1}{\rule{0.8em}{0.8em}}}

\definecolor{categorybg}{RGB}{248, 240, 240}

\newcommand{\cell}[2]{\cellcolor{#1}#2}
\newcommand{\best}[2]{\cellcolor{#1}\textbf{#2}}

\begin{table*}[t]
  \centering
  \setlength{\tabcolsep}{1.6pt} 
  \caption{Comprehensive quantitative evaluation results. The metrics are categorized into four dimensions: \textbf{Visual Quality} (Dov.: Dover Score, Mus.: MusIQ Score, V.A.C: Visual Attribute Consistency, S.C.: Style Consistency); \textbf{Story Video Alignment} (VQA: VQAScore, Det.: Detection \& Count Score, S.P.A: Shot Perspective Alignment, S.S.P: State Shift \& Persistence, S.V.C: Story Video Consistency); \textbf{Video Consistency} (Fac.: Face Consistency, Cha.: Character Consistency, Bac.: Background Consistency, Clo.: Clothes \& Color Consistency, Siz.: Relative Size Consistency); and \textbf{Motion Quality} (A.R.: Action Recognition, A.S.: Action Strength, Cam.: Camera Control, Phy.P: Physical Plausibility, Phy.I: Physical Interaction Accuracy). The background colors 
\protect\rankbox{rankA}
\protect\rankbox{rankB}
\protect\rankbox{rankC}
\protect\rankbox{rankD}
\protect\rankbox{rankE}
\protect\rankbox{rankF}
\protect\rankbox{rankG}
\protect\rankbox{rankH}
represent the best to the 8th best models, respectively. In cases of ties, the median color is adopted.}
  \label{tab:full_evaluation}
  
  \resizebox{\textwidth}{!}{
    {\fontsize{13}{16}\selectfont
      \begin{tabular}{l cccc ccccc ccccc ccccc}
        \toprule
        & \multicolumn{4}{c}{\textbf{Visual Quality}} & \multicolumn{5}{c}{\textbf{Story Video Alignment}} & \multicolumn{5}{c}{\textbf{Video Consistency}} & \multicolumn{5}{c}{\textbf{Motion Quality}} \\
        \cmidrule(lr){2-5} \cmidrule(lr){6-10} \cmidrule(lr){11-15} \cmidrule(lr){16-20}
        \textbf{Model} & \textbf{Dov.} & \textbf{Mus.} & \textbf{V.A.C} & \textbf{S.C.} & \textbf{VQA} & \textbf{Det.} & \textbf{S.P.A} & \textbf{S.S.P} & \textbf{S.V.C} & \textbf{Fac.} & \textbf{Cha.} & \textbf{Bac.} & \textbf{Clo.} & \textbf{Siz.} & \textbf{A.R.} & \textbf{A.S.} & \textbf{Cam.} & \textbf{Phy.P} & \textbf{Phy.I} \\
        \midrule
        
        \multicolumn{20}{c}{\cellcolor{categorybg}\textbf{Storyboard-Driven Synthesis}} \\
        AniSora3.2 & \cell{rankC}{0.68} & \cell{rankC}{0.21} & 81.50 & \cell{rankH}{0.96} & \cell{rankG}{81.72} & \cell{rankG}{83.02} & \cell{rankC}{4.20} & \cell{rankE}{3.93} & \cell{rankD}{81.41} & 87.05 & 85.38 & \cell{rankG}{0.53} & \cell{rankB}{3.63} & 0.98 & 23.57 & 17.41 & 3.57 & \cell{rankF}{3.41} & \cell{rankG}{2.41} \\
        CogVideoX1.5-5B-I2V & \cell{rankE}{0.67} & 0.19 & 74.21 & 0.95 & \cell{rankD}{82.64} & \cell{rankE}{86.01} & \cell{rankD}{4.19} & \cell{rankC}{4.05} & 81.20 & 87.73 & \cell{rankF}{86.03} & 0.52 & \cell{rankB}{3.63} & 0.98 & 23.16 & 16.00 & 3.45 & 3.14 & 2.29 \\
        HunyuanVideo-I2V & 0.58 & \cell{rankG}{0.20} & 74.86 & 0.95 & \cell{rankH}{81.30} & 77.77 & \cell{rankF}{4.13} & \cell{rankH}{3.72} & \cell{rankH}{81.34} & \cell{rankB}{89.14} & \cell{rankD}{86.08} & 0.52 & 3.36 & \cell{rankG}{0.99} & 24.90 & 31.44 & \cell{rankG}{3.58} & 3.11 & \cell{rankF}{2.43} \\
        LTXV-13B-0.9.8 & \cell{rankE}{0.67} & 0.19 & 74.97 & 0.94 & \cell{rankF}{81.87} & 76.93 & \cell{rankH}{4.07} & \cell{rankG}{3.83} & \cell{rankB}{81.69} & 88.25 & \cell{rankC}{86.12} & \cell{rankE}{0.55} & \cell{rankE}{3.61} & \cell{rankG}{0.99} & 24.84 & 35.22 & 3.52 & 2.76 & \cell{rankH}{2.34} \\
        S.A.+Wan2.2-I2V & 0.57 & \cell{rankG}{0.20} & 86.39 & \cell{rankH}{0.96} & 71.95 & 65.28 & 3.54 & 2.81 & 80.55 & \cell{rankC}{88.97} & 81.68 & 0.43 & 0.84 & \cell{rankG}{0.99} & \cell{rankH}{29.76} & \cell{rankD}{56.76} & \cell{rankB}{3.67} & 2.58 & 1.63 \\
        S.D.+Wan2.2-I2V & 0.63 & \cell{rankG}{0.20} & 75.53 & 0.94 & 61.88 & 55.83 & 3.60 & 2.33 & 79.93 & \cell{rankF}{88.84} & 82.12 & \cell{rankB}{0.56} & 0.96 & \cell{rankB}{1.00} & \cell{rankG}{30.59} & \cell{rankA}{73.80} & \cell{rankF}{3.59} & 2.47 & 1.24 \\
        Self-Forcing & \cell{rankA}{0.71} & 0.16 & 83.28 & 0.96 & \cell{rankB}{85.10} & 69.60 & 3.99 & 3.28 & 81.11 & \cell{rankG}{88.66} & 83.65 & \cell{rankB}{0.56} & 2.10 & 0.97 & \cell{rankE}{32.48} & \cell{rankE}{52.58} & \cell{rankG}{3.57} & 3.15 & 1.93 \\
        StoryAdapter & 0.47 & 0.20 & \cell{rankA}{93.59} & \cell{rankD}{0.98} & 67.25 & 54.94 & 3.47 & 1.48 & 80.39 & \cell{rankH}{88.38} & 79.41 & 0.39 & 0.78 & \cell{rankB}{1.00} & \cell{rankC}{33.16} & 0.47 & \cell{rankC}{3.65} & 0.77 & 0.79 \\
        StoryDiffusion & 0.54 & 0.20 & \cell{rankH}{87.73} & 0.96 & 67.51 & 54.68 & 3.66 & 1.72 & 80.52 & \cell{rankA}{89.95} & 81.92 & 0.51 & 1.05 & \cell{rankB}{1.00} & \cell{rankA}{36.33} & 0.52 & 3.41 & 0.76 & 0.87 \\
        StoryGen & 0.25 & 0.19 & \cell{rankG}{88.38} & 0.96 & 32.82 & 14.16 & 3.27 & 1.28 & 76.60 & 79.70 & 82.05 & 0.31 & 0.37 & 0.93 & 25.91 & 0.66 & 3.48 & 0.26 & 0.33 \\
        Wan2.2-I2V & \cell{rankG}{0.65} & 0.20 & 79.36 & 0.92 & \cell{rankC}{83.22} & \cell{rankA}{90.90} & \cell{rankG}{4.07} & \cell{rankC}{4.05} & \cell{rankC}{81.48} & \cell{rankE}{88.85} & 85.10 & \cell{rankB}{0.56} & \cell{rankD}{3.62} & 0.97 & 25.16 & 26.64 & 3.53 & \cell{rankE}{3.44} & 2.13 \\
        Wan2.2-T2V & 0.65 & 0.16 & 82.58 & 0.94 & \cell{rankA}{86.22} & \cell{rankC}{88.66} & 3.99 & 3.43 & \cell{rankE}{81.40} & \cell{rankD}{88.94} & 84.27 & \cell{rankB}{0.56} & 2.04 & 0.97 & 27.32 & \cell{rankH}{46.90} & \cell{rankD}{3.60} & \cell{rankA}{3.86} & \cell{rankA}{2.87} \\
        Wan2.2-TI2V & \cell{rankC}{0.68} & 0.20 & 78.48 & 0.95 & 78.71 & \cell{rankH}{78.84} & \cell{rankB}{4.21} & 3.61 & 81.33 & 88.34 & \cell{rankG}{85.97} & \cell{rankF}{0.54} & \cell{rankG}{3.56} & \cell{rankG}{0.99} & 25.76 & \cell{rankF}{50.26} & \cell{rankD}{3.60} & \cell{rankG}{3.26} & 2.31 \\
        
        \multicolumn{20}{c}{\cellcolor{categorybg}\textbf{Agent-Based Frameworks}} \\
        AniMaker & \cell{rankB}{0.69} & \cell{rankC}{0.21} & 81.90 & 0.95 & \cell{rankE}{82.13} & \cell{rankE}{86.01} & 4.05 & \cell{rankF}{3.88} & \cell{rankG}{81.39} & 87.73 & \cell{rankE}{86.07} & \cell{rankG}{0.53} & \cell{rankH}{3.41} & 0.95 & 27.07 & \cell{rankG}{49.41} & 3.45 & \cell{rankH}{3.24} & \cell{rankE}{2.52} \\
        MM-StoryAgent & 0.45 & \cell{rankC}{0.21} & \cell{rankF}{90.13} & \cell{rankD}{0.98} & 34.27 & - & - & - & 79.69 & - & - & 0.44 & - & - & 14.22 & 0.69 & - & 0.76 & 0.79 \\
        MovieAgent & 0.61 & 0.18 & 72.50 & \cell{rankH}{0.96} & 47.84 & 45.71 & 3.54 & 1.29 & 79.14 & - & - & 0.45 & 0.77 & 0.97 & 24.90 & 28.43 & 2.89 & 2.54 & 1.68 \\
        VideoGen-of-Thought & 0.56 & \cell{rankC}{0.21} & \cell{rankB}{91.75} & \cell{rankA}{1.00} & 60.11 & 32.07 & 3.56 & 1.68 & 80.51 & 83.58 & \cell{rankA}{87.70} & 0.42 & 1.22 & \cell{rankB}{1.00} & 25.11 & \cell{rankB}{63.36} & 3.55 & 0.71 & 0.59 \\
    
        \multicolumn{20}{c}{\cellcolor{categorybg}\textbf{Continuous Generation Techniques}} \\
        LongLive & 0.60 & 0.16 & \cell{rankC}{91.66} & \cell{rankD}{0.98} & 67.75 & - & - & - & 77.10 & - & - & - & - & - & \cell{rankF}{30.62} & \cell{rankC}{57.36} & - & \cell{rankD}{3.49} & \cell{rankD}{2.56} \\
        
        \multicolumn{20}{c}{\cellcolor{categorybg}\textbf{Commercial Solutions}} \\
        Sora2 & \cell{rankG}{0.65} & 0.19 & \cell{rankE}{90.57} & \cell{rankD}{0.98} & 79.15 & \cell{rankD}{88.58} & \cell{rankA}{4.23} & \cell{rankB}{4.09} & \cell{rankE}{81.40} & 87.66 & \cell{rankB}{86.43} & 0.46 & \cell{rankF}{3.60} & \cell{rankG}{0.99} & \cell{rankD}{33.08} & 30.86 & 3.56 & \cell{rankB}{3.74} & \cell{rankC}{2.64} \\
        Veo3.1 & 0.64 & \cell{rankC}{0.21} & \cell{rankD}{91.04} & \cell{rankD}{0.98} & 77.91 & \cell{rankB}{90.67} & \cell{rankE}{4.18} & \cell{rankA}{4.10} & \cell{rankA}{82.24} & 86.64 & \cell{rankH}{85.77} & 0.51 & \cell{rankA}{3.66} & \cell{rankG}{0.99} & \cell{rankB}{33.27} & 34.80 & \cell{rankA}{3.68} & \cell{rankC}{3.67} & \cell{rankB}{2.78} \\
        \bottomrule
      \end{tabular}
    }
  }
\end{table*}

\subsection{Settings}
We evaluate a comprehensive range of methods on MSVBench across four distinct paradigms: (1) Storyboard-Driven Synthesis: We assess StoryGen \cite{liu2024intelligent}, StoryDiffusion \cite{zhou2024storydiffusion}, and StoryAdapter \cite{mao2024story}, alongside decoupled I2V pipelines (Wan series \cite{wan2025wan}, HunyuanVideo \cite{kong2024hunyuanvideo}, AniSora \cite{jiang2024anisora}) using \texttt{GPT-Image-1} keyframes. (2) Agent-Based Frameworks: We benchmark systems including MovieAgent \cite{wu2025automated}, MM-StoryAgent \cite{xu2025mm}, VideoGen-of-Thought \cite{zheng2024videogen}, and AniMaker \cite{shi2025animaker}. (3) Continuous Generation Techniques: We evaluate the autoregressive method LongLive \cite{yang2025longlive}. (4) Commercial Solutions: Models like Sora \cite{sora2024} and Veo 3 \cite{google_veo3_2025} are included.

\subsection{Main Results}
Table~\ref{tab:full_evaluation} details the performance of 20 multi-shot video generation methods across four dimensions. Commercial models (Sora2, Veo3.1) currently define the state-of-the-art, exhibiting superior robustness in Story Video Alignment and Motion Quality. However, the gap with open-source systems is rapidly narrowing. The Wan2.2 family emerges as the premier open-source contender; remarkably, Wan2.2-I2V achieves parity with commercial models in Video Consistency, while Wan2.2-T2V demonstrates highly competitive Motion Quality. Within Agent-Based frameworks, AniMaker distinguishes itself through its well-balanced performance profile across all evaluation dimensions.

\subsection{Critical Insights} 
Synthesizing the quantitative results of evaluated methods and case studies of evaluation metrics—representative examples of which are visualized in Appendix~\ref{sec:QuantitativeResults} and ~\ref{sec:CaseStudy}—we move beyond surface-level rankings to uncover fundamental architectural constraints preventing current models from functioning as true world models. Three structural impediments are identified:

\noindent\textbf{Fragmented Generation rather than Holistic Modeling.} 
While current models excel at single-shot interpretation—evidenced by high Story Video Alignment (S.V.A)—they largely fail to \textbf{model} the generated content. This deficiency manifests at two scales: locally, low Physical Interaction Accuracy (Phy.I) scores (< 3.0 / 5 even for Sora2 and Veo3.1) indicate an inability to model immediate causal dynamics such as collisions; globally, the degradation of Character Consistency (Cha.) and Clothing (Clo.) across multi-shot narratives reveals a failure to maintain a coherent character-level model that preserves identity and attributes. These failures suggest that current models function primarily as local visual interpolators rather than holistic world models, as they lack the capacity to maintain internal representation that governs physical laws and semantic consistency.

\noindent\textbf{Trade-offs in Cross-Dimensional Model Capabilities.}
An inherent conflict exists between dynamic intensity and content preservation in current architectures. This tension manifests in two dimensions: First, in object dynamics, increasing Action Strength (A.S.) compromises Physical Interaction Accuracy (Phy.I), exemplified by S.D.+Wan2.2-I2V's exceptional A.S. (73.80 / 100) versus negligible Phy.I (1.24 / 5). This disparity confirms that extreme motion generation often distorts object structures, undermining the preservation of valid physical states. Second, in viewpoint dynamics, aggressive Camera Control (Cam.) disrupts Character Consistency (Cha.), as models prioritize rapid view shifts over semantic identity. These failures indicate that motion and stability mechanisms remain deeply entangled; thus, future designs that explicitly decouple motion generation from content preservation hold significant promise.

\noindent\textbf{Reference Images as a Double-Edged Sword.}
While reference images provide dense visual guidance, enabling open-source models like Wan2.2-I2V to achieve Video Consistency comparable to commercial giants, they simultaneously act as a rigid constraint. Specifically, the static image \textbf{locks} the initial state but fails to convey depth or kinematic potential. Consequently, Wan2.2-T2V, unburdened by this 2D anchor, outperforms its I2V counterpart in Physical Plausibility (3.86 vs. 3.44). This performance gap highlights the intrinsic limitations of using 2D pixels as the sole conditioning signal; to resolve this, future generation paradigms must incorporate more comprehensive geometric inputs, such as 3D meshes or depth priors.

\subsection{Human Preference Alignment}

\begin{table}[t]
  \centering
  \caption{Human evaluation results. We report Mean Opinion Scores across four dimensions: \textbf{V.Q.} (Visual Quality), \textbf{S.V.A.} (Story Video Alignment), \textbf{V.C.} (Video Consistency), and \textbf{M.Q.} (Motion Quality). The models are ranked by the overall average (\textbf{Avg.}). We highlight the \textbf{best}, \underline{second-best}, and \textit{third-best} results.}
  \label{tab:human_ratings}
  
  \resizebox{\columnwidth}{!}{
    \begin{tabular}{lccccc}
        \toprule
        \textbf{Model} & \textbf{V.Q.} & \textbf{S.V.A.} & \textbf{V.C.} & \textbf{M.Q.} & \textbf{Avg.} \\
        \midrule
        Veo3.1 & \textbf{4.29} & \textbf{4.51} & \textbf{3.68} & \underline{3.97} & \textbf{4.11} \\
        Sora2 & \underline{4.22} & \underline{4.45} & \underline{3.56} & \textbf{4.01} & \underline{4.06} \\
        Anisora3.2 & \textit{3.98} & 3.40 & 3.03 & 3.32 & \textit{3.43} \\
        Wan2.2-I2V & 3.70 & 3.48 & 2.86 & \textit{3.40} & 3.36 \\
        LTXV-13B-0.9.8 & 3.90 & 3.27 & 3.06 & 3.22 & 3.36 \\
        AniMaker & 3.04 & 3.42 & \textit{3.14} & 3.08 & 3.17 \\
        Wan2.2-T2V & 3.22 & \textit{3.63} & 2.15 & 3.34 & 3.08 \\
        Wan2.2-TI2V & 3.04 & 3.15 & 2.67 & 2.83 & 2.92 \\
        HunyuanVideo-I2V & 2.71 & 2.92 & 2.82 & 2.58 & 2.76 \\
        Self-Forcing & 2.96 & 2.77 & 2.36 & 2.65 & 2.69 \\
        CogVideoX1.5-5B-I2V & 2.01 & 2.81 & 2.48 & 2.26 & 2.39 \\
        VideoGen-of-Thought & 2.97 & 1.67 & 2.30 & 1.33 & 2.07 \\
        S.D.+Wan2.2-I2V & 2.76 & 1.76 & 1.68 & 1.95 & 2.04 \\
        LongLive & 2.67 & 1.31 & 2.69 & 1.51 & 2.04 \\
        S.A.+Wan2.2-I2V & 2.92 & 1.74 & 1.68 & 1.77 & 2.03 \\
        StoryDiffusion & 2.96 & 1.60 & 1.80 & 1.02 & 1.85 \\
        StoryAdapter & 3.00 & 1.58 & 1.77 & 1.00 & 1.84 \\
        MM-StoryAgent & 2.26 & 1.44 & 1.38 & 1.00 & 1.52 \\
        MovieAgent & 1.89 & 1.12 & 1.39 & 1.03 & 1.36 \\
        StoryGen & 1.03 & 1.01 & 1.02 & 1.00 & 1.01 \\
        \bottomrule
    \end{tabular}
  }
\end{table}

\begin{table}[t]
  \centering
  \caption{Correlation Analysis. Comparison of Spearman’s ($\rho$) and Kendall’s ($\tau$) correlation coefficients between objective metrics and human judgments across different dimensions. We compare MSVBench with current evaluation metrics including VBench~\cite{huang2024vbench}, EvalCrafter~\cite{liu2024evalcrafter} and ViStoryBench~\cite{zhuang2025vistorybench}. We highlight the \textbf{best} and \underline{second-best} results.}
  \label{tab:correlation_comparison}
  \resizebox{\columnwidth}{!}{
    \begin{tabular}{cl|cc}
      \toprule
      \multirow{2}{*}{\textbf{Aspects}} & \multirow{2}{*}{\textbf{Methods}} & \textbf{Spearman's} & \textbf{Kendall's} \\
       & & $\rho$ (\%) & $\tau$ (\%) \\
      \midrule
      \multirow{4}{*}{\shortstack{Visual\\Quality}} 
        & Aesthetic-Quality (VBench) & \underline{47.0} & \underline{39.5} \\
        & Imaging-Quality (VBench) & 35.1 & 27.8 \\
        & ViStoryBench & 27.2 & 15.4 \\
        & \textbf{MSVBench (Ours)} & \textbf{60.4} & \textbf{47.9} \\
      \midrule
      \multirow{3}{*}{\shortstack{Video Story\\Alignment}} 
        & Clip-Score (EvalCrafter) & 42.5 & 35.2 \\
        & BLIP-BLEU (EvalCrafter) & 17.5 & 11.8 \\
        & ViStoryBench & \underline{80.5} & \underline{62.1} \\
        & \textbf{MSVBench (Ours)} & \textbf{94.5} & \textbf{83.9} \\
      \midrule
      \multirow{5}{*}{\shortstack{Video\\Consistency}} 
        & Subject Consistency (VBench) & 54.3 & 43.2 \\
        & Background Consistency (VBench) & 44.3 & 35.5 \\
        & CLIP-Temp (EvalCrafter) & 1.7 & 1.4 \\
        & ViStoryBench & \textbf{66.5} & \textbf{48.7} \\
        & \textbf{MSVBench (Ours)} & \underline{60.0} & \underline{44.1} \\
      \midrule
      \multirow{3}{*}{\shortstack{Motion\\Quality}} 
        & Motion-Smoothness (VBench) & 51.1 & \underline{43.0} \\
        & Dynamic-Degree (VBench) & \underline{52.9} & 36.6 \\
        & \textbf{MSVBench (Ours)} & \textbf{64.9} & \textbf{47.1} \\
      \bottomrule
      \toprule
      \multirow{3}{*}{\textbf{Overall}} 
        & VBench & 58.5 & 50.4 \\
        & EvalCrafter & 26.3 & 14.4 \\
        & ViStoryBench & \underline{83.6} & \underline{66.0} \\
        & \textbf{MSVBench (Ours)} & \textbf{94.4} & \textbf{83.6} \\
      \bottomrule
    \end{tabular}
  }
\end{table}

\begin{table}[t]
\centering
\caption{Comparison of overall human alignment (Spearman's $\rho$ and Kendall's $\tau$) between the lightweight Qwen3-VL models and proprietary Gemini models. We highlight the \textbf{best} and \underline{second-best} results.}
\label{tab:qwen_alignment}
\resizebox{0.9\columnwidth}{!}{%
\begin{tabular}{l|cc}
\toprule
\multirow{2}{*}{\textbf{Model}} & \multicolumn{2}{c}{\textbf{Overall Correlation (\%)}} \\
 & Spearman's $\rho$ & Kendall's $\tau$ \\
\midrule
\textit{Commercial Models} & & \\
Gemini-2.5-Flash & 79.2 & 62.8 \\
Gemini-2.5-Pro & \textbf{85.7} & \textbf{70.2} \\
\midrule
\textit{Ours (Lightweight)} & & \\
Qwen3-VL-4B (Base) & 79.9 & 59.6 \\
Qwen3-VL-4B (RL) & \underline{83.6} & \underline{66.3} \\
\bottomrule
\end{tabular}%
}
\end{table}

\paragraph{Human Evaluation Results}
We engage a group of human annotators to assess the generated videos across four key dimensions: Visual Quality (\textbf{V.Q.}), Story Video Alignment (\textbf{S.V.A.}), Video Consistency (\textbf{V.C.}), and Motion Quality (\textbf{M.Q.}). The detailed human evaluation setup, including annotator guidelines, is described in Appendix~\ref{sec:HumanRatingDetails}. The results of the human evaluation are presented in Table~\ref{tab:human_ratings}. 

\paragraph{Alignment with Human Perception}
We employ Spearman’s ($\rho$) and Kendall’s ($\tau$) rank correlations to evaluate the consistency between benchmarks and human ratings. The detailed methodology is provided in Appendix~\ref{sec:CorrelationCoefficientDetails}. Table~\ref{tab:correlation_comparison} presents MSVBench achieves an overall $\rho$ of \textbf{94.4\%} and $\tau$ of \textbf{83.6\%}, significantly outperforming state-of-the-art baselines like VBench ($\rho=58.5\%$, $\tau=50.4\%$) and ViStoryBench ($\rho=83.6\%$, $\tau=66.0\%$). Notably, these overall correlations surpass the scores of individual dimensions across both metrics (approx. 60\% for $\rho$ and 45\% for $\tau$). This validates that the proposed metrics synergistically complement each other to accurately capture the holistic nature of human judgments.

\subsection{From Benchmark to Supervisor}
To verify the effectiveness of MSVBench as a source of high-quality supervision signals, we implement a data construction pipeline that transforms raw evaluation records into high-quality instruction tuning data. Using a subset of 15 stories, we synthesize over 1,000 samples and employee a Qwen3-VL-4B backbone trained with Group Relative Policy Optimization (GRPO) for a single epoch. As shown in Table~\ref{tab:qwen_alignment}, the model trained on our dataset achieves Spearman's $\rho$ of \textbf{83.6\%} and Kendall's $\tau$ of \textbf{66.3\%}, effectively outperforming Gemini-2.5-Flash. Beyond these aggregate metrics, the data successfully transfers the fine-grained discriminative criteria of human annotators to the model: the scoring distribution shifts from a ``conservative'' clustering around the median (score 3) to a confident coverage of the full 1--5 range. This dual improvement—high alignment accuracy combined with human-like variance—confirms MSVBench's capability as a robust supervisor, effectively transferring comprehensive preference patterns to downstream models.

\section{Conclusion}
We introduce MSVBench, the first unified framework for evaluating multi-shot video generation. By synergizing specialized expert models for perceptual precision with LMMs for semantic reasoning, our hybrid protocol effectively bridges the gap between low-level visual fidelity and high-level narrative consistency. Extensive validation confirms that MSVBench achieves state-of-the-art human alignment, significantly outperforming all existing video generation benchmarks. Beyond rigorously benchmarking current systems and identifying their limitations as world models, MSVBench serves as an automated pipeline for generating high-quality supervision data, enabling a lightweight Qwen3-VL-4B model trained on its reasoning traces to achieve more human-aligned video evaluation.


\section*{Limitations}
Despite the comprehensive nature of MSVBench, several limitations remain to be addressed in future work:
\begin{itemize}
    \item \textbf{Lack of Audio-Visual Metrics:} The current framework is predominantly centered on visual evaluation. It does not yet incorporate metrics for assessing audio-visual synchronization or the quality of generated audio—both of which are pivotal components for achieving immersive video generation.
    \item \textbf{Limited Scale of Story Data:} The benchmark currently comprises a relatively small number of stories. This scarcity limits the volume of training data available for the data reconstruction pipeline used to train our lightweight evaluator. Consequently, the potential effectiveness of the automated evaluator may not be fully realized due to the data-constrained training regime.
    \item \textbf{Challenges with Continuous Generation Models:} For continuous generation approaches that synthesize long videos without explicit internal shot segmentation, our evaluation pipeline faces alignment difficulties. The absence of discrete shot boundaries hinders the accurate mapping of corresponding text and reference image prompts, thereby rendering a specific subset of shot-level metrics inapplicable or difficult to compute.
\end{itemize}


\clearpage

\bibliography{custom}

\clearpage

\appendix

\section{MSVBench Prompt}
\label{sec:MSVBenchPrompt}
This section details the comprehensive set of prompts tailored for the LMM evaluator, systematically structured to evaluate performance across key dimensions: Story Video Alignment, Video Consistency, and Motion Quality. 

\begin{figure}[h!]
    \centering
    \includegraphics[width=\columnwidth]{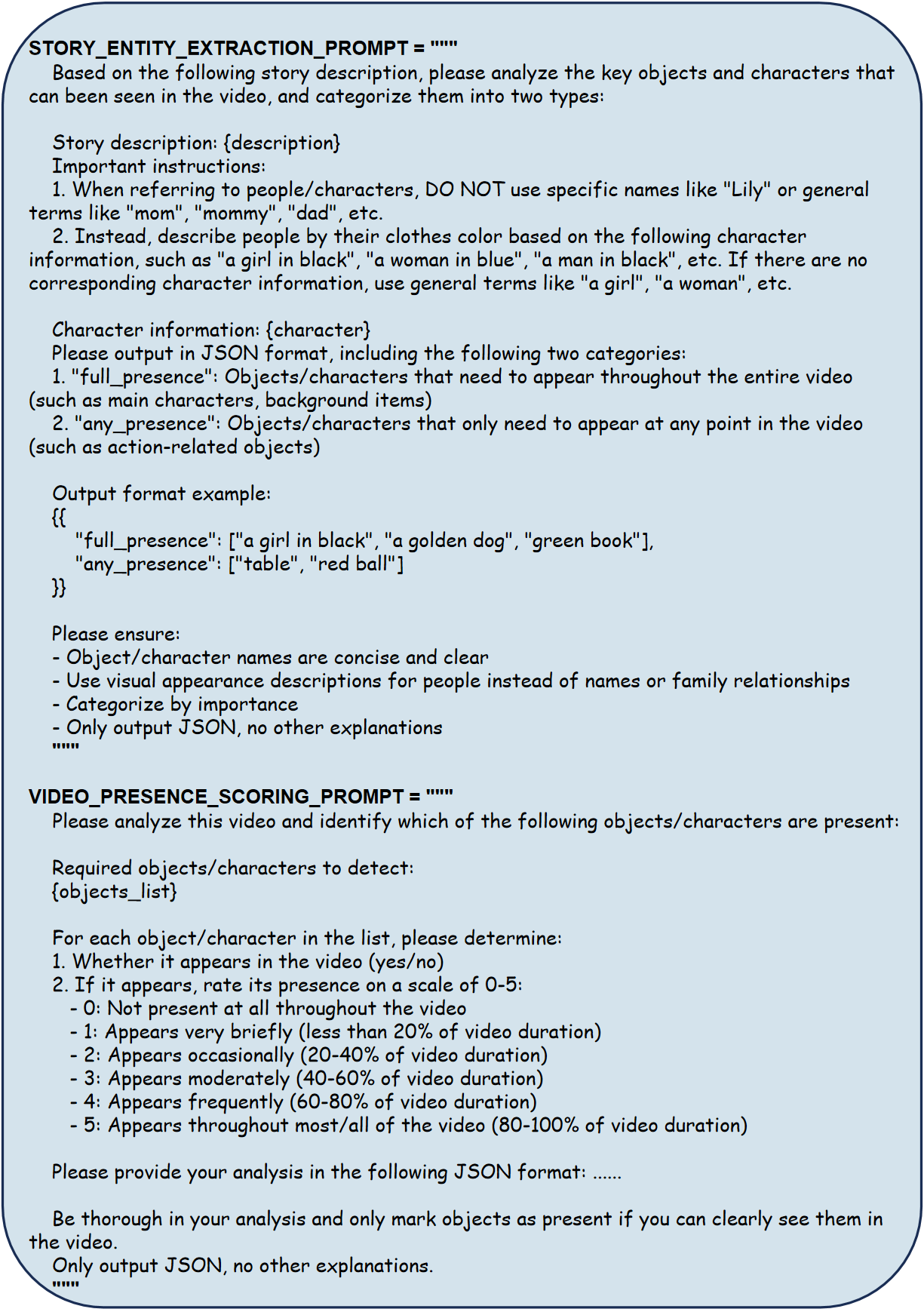}
    \caption{Prompt for Detect \& Count Score}
    \label{fig:Detect & Count Score}
\end{figure}

\begin{figure}
    \centering
    \includegraphics[width=\columnwidth]{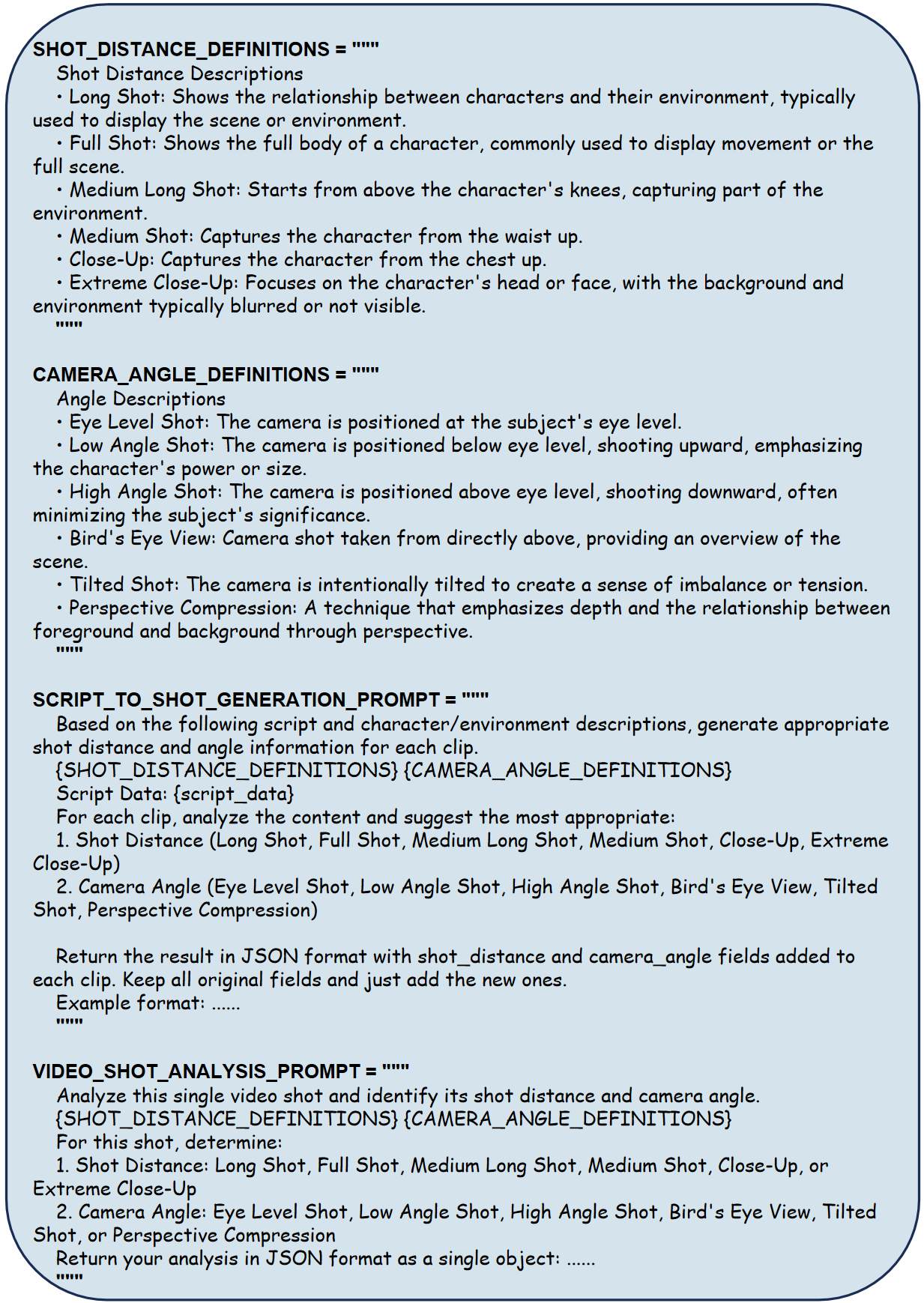}
    \caption{Prompt for Shot Perspective Alignment}
    \label{fig:Shot Perspective Alignment}
\end{figure}

\begin{figure}
    \centering
    \includegraphics[width=\columnwidth]{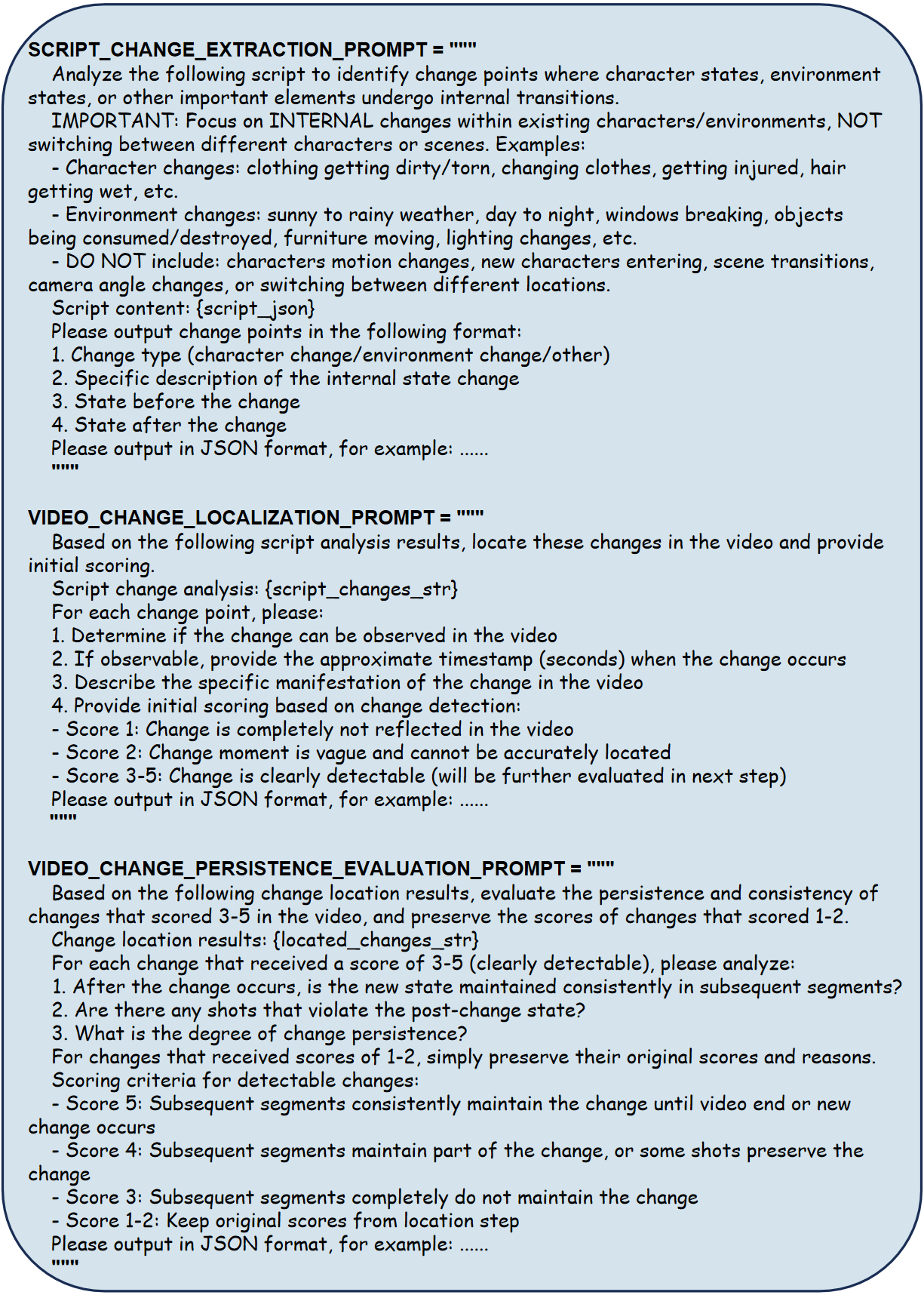}
    \caption{Prompt for State Shift \& Persistence}
    \label{fig:State Shift & Persistence}
\end{figure}

\begin{figure}
    \centering
    \includegraphics[width=\columnwidth]{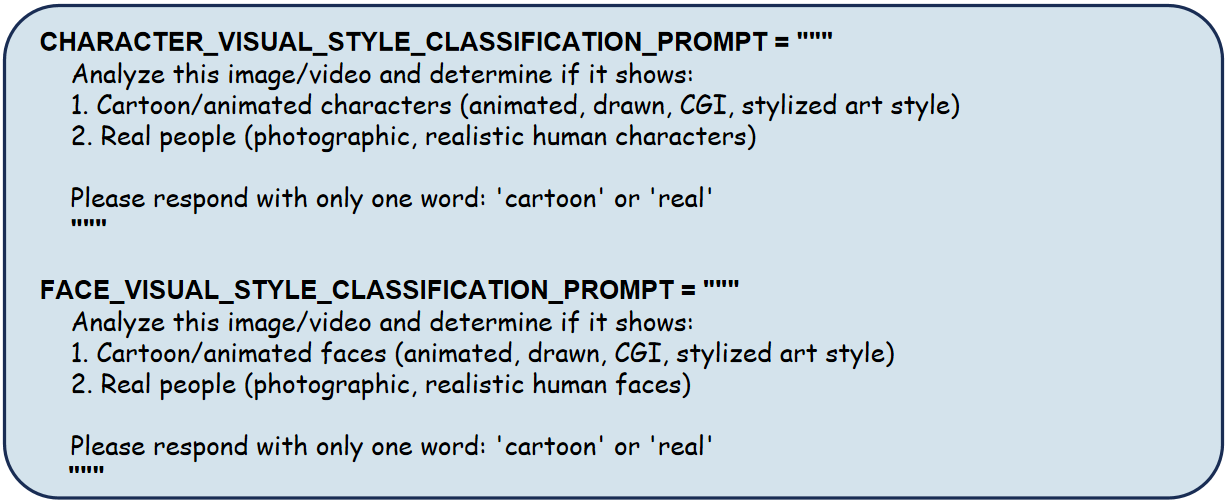}
    \caption{Prompt for Character \& Face Consistency}
    \label{fig:Character & Face Consistency}
\end{figure}

\begin{figure}
    \centering
    \includegraphics[width=\columnwidth]{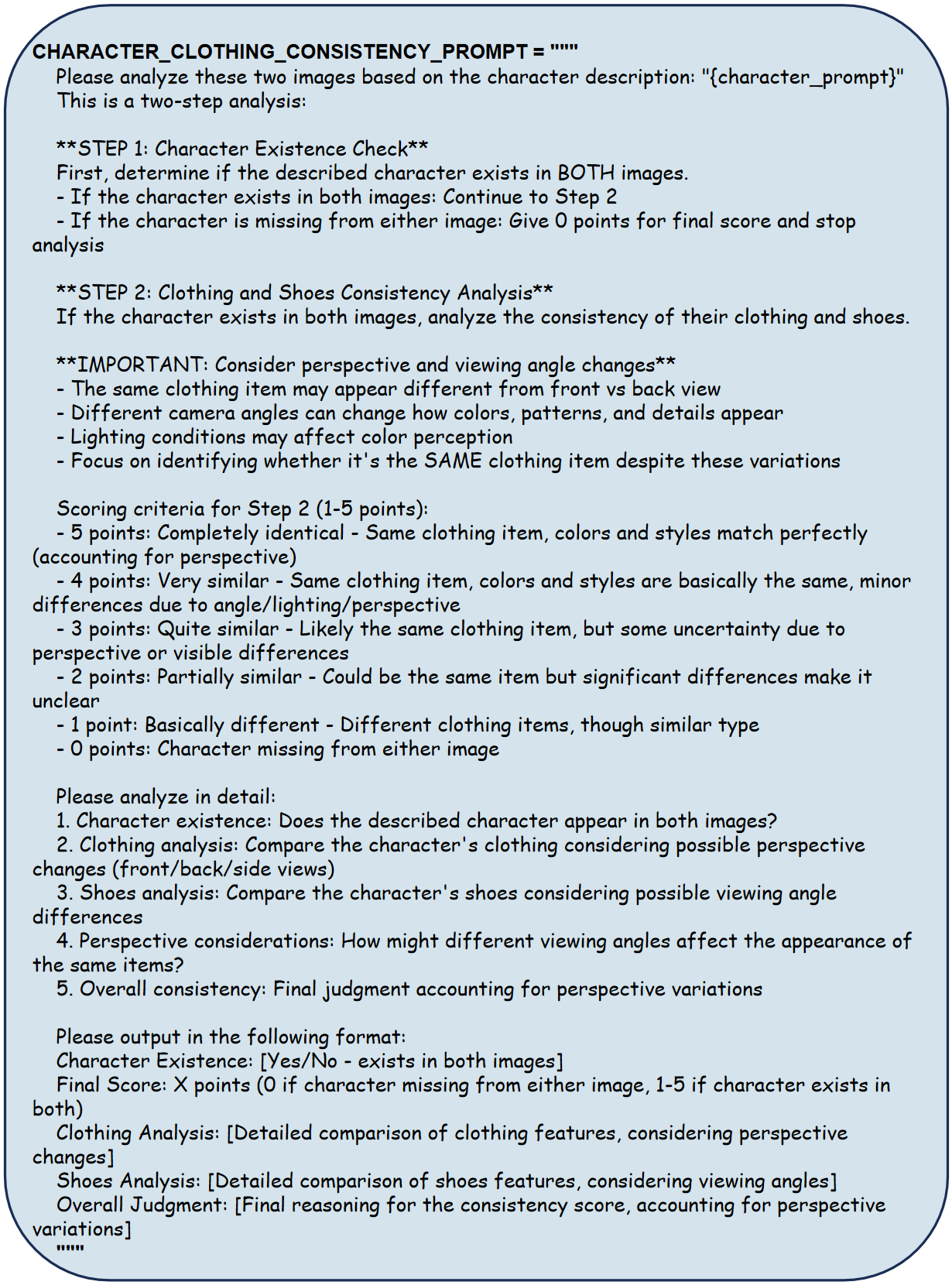}
    \caption{Prompt for Clothes \& Color Consistency}
    \label{fig:Clothes & Color Consistency}
\end{figure}

\begin{figure}
    \centering
    \includegraphics[width=\columnwidth]{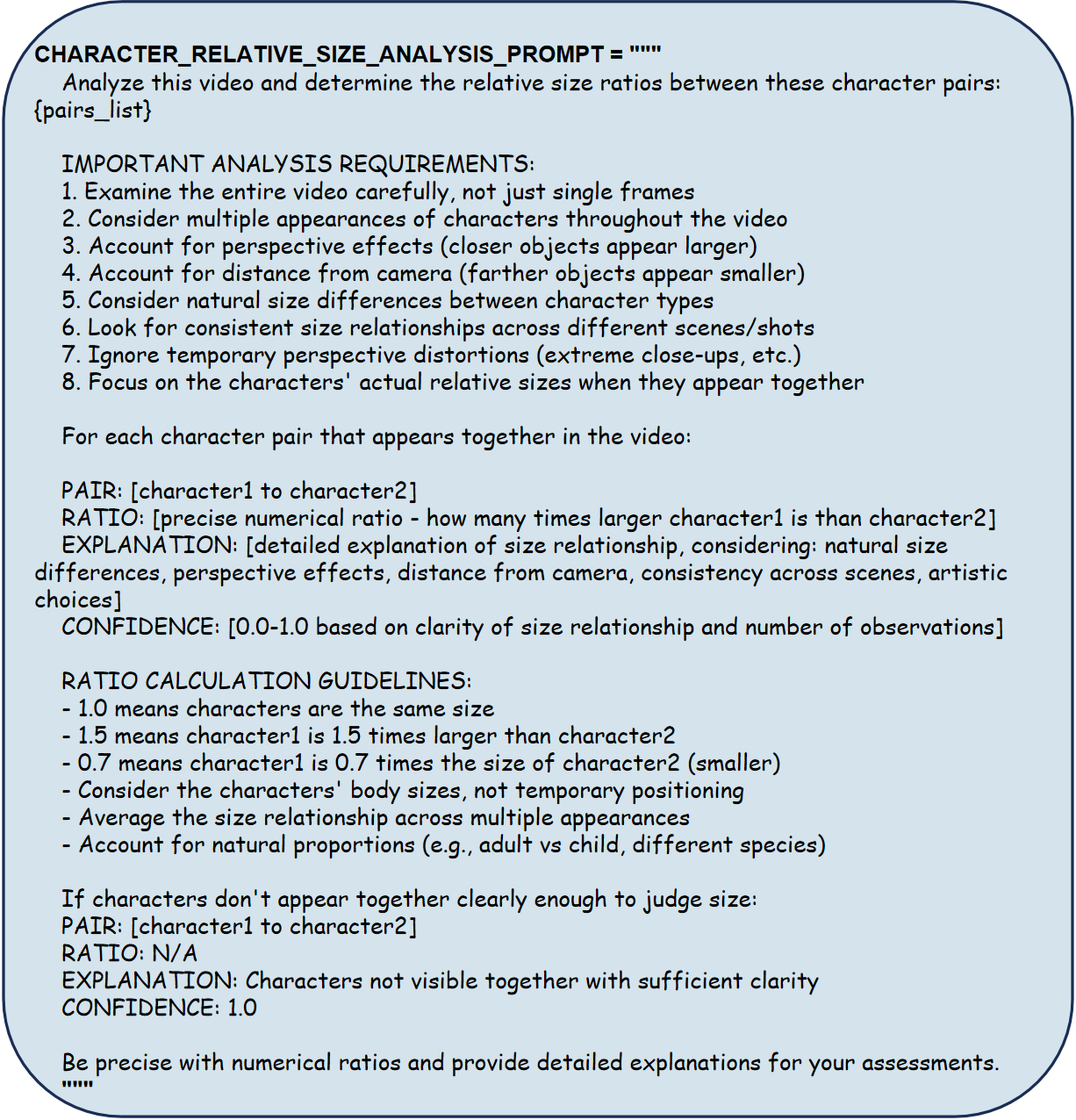}
    \caption{Prompt for Relative Size Consistency}
    \label{fig:Relative Size Consistency}
\end{figure}

\begin{figure}
    \centering
    \includegraphics[width=\columnwidth]{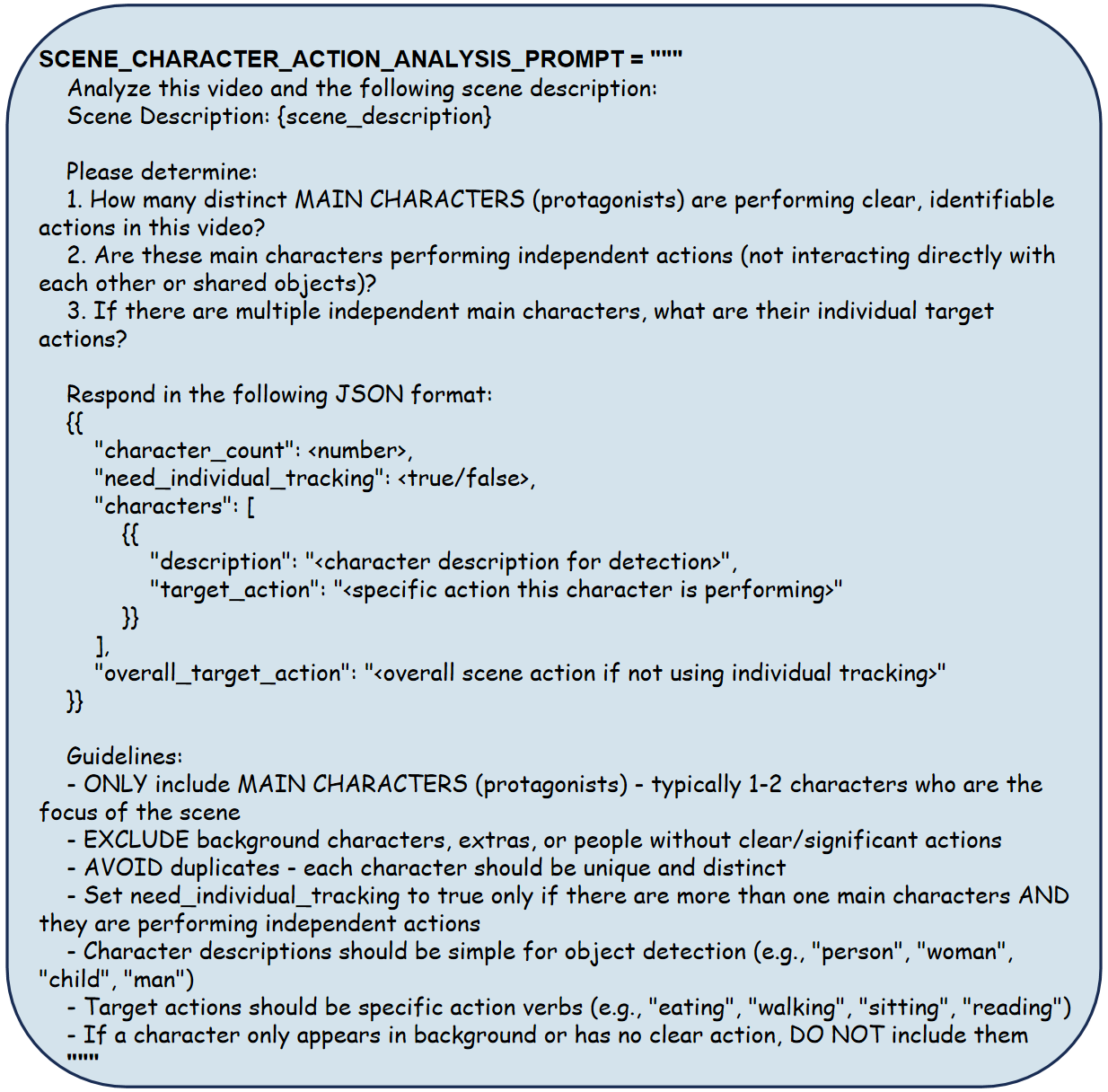}
    \caption{Prompt for Action Recognition}
    \label{fig:Action Recognition}
\end{figure}

\begin{figure}
    \centering
    \includegraphics[width=\columnwidth]{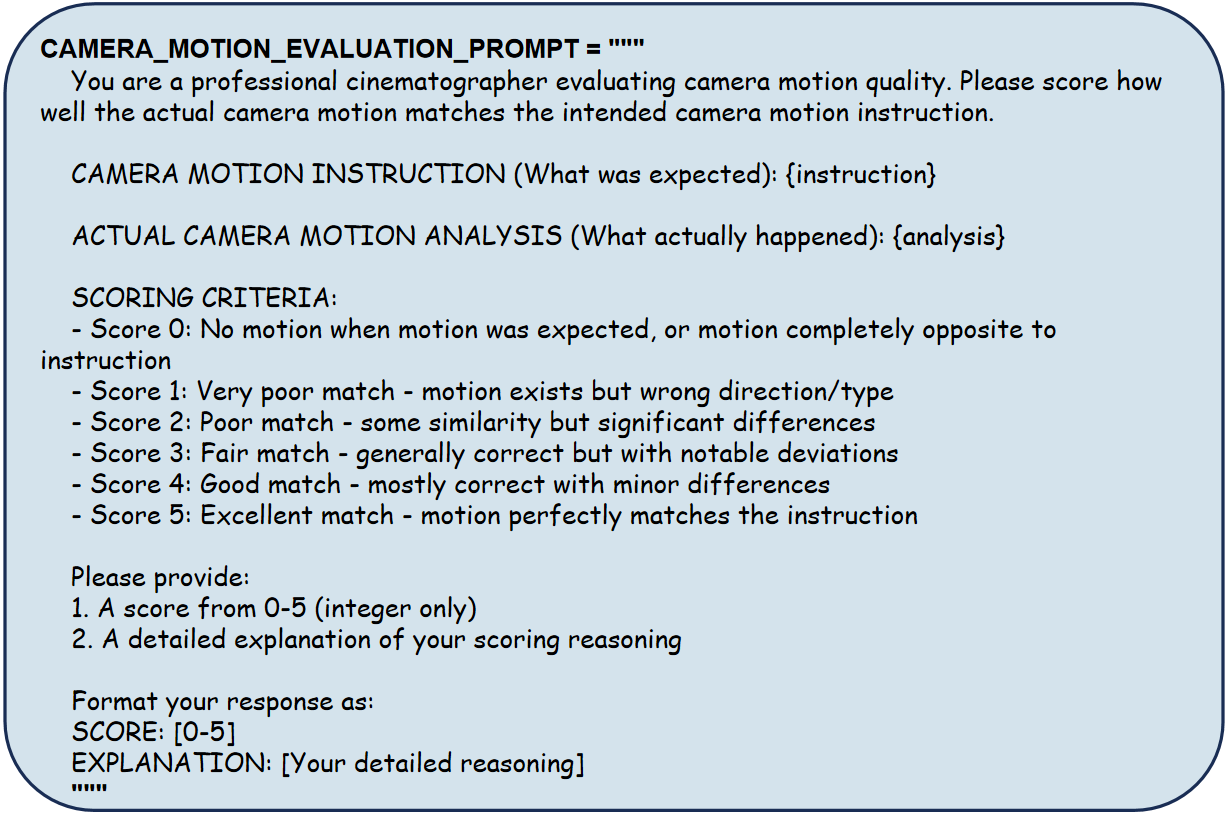}
    \caption{Prompt for Camera Control}
    \label{fig:Camera Control}
\end{figure}

\begin{figure}
    \centering
    \includegraphics[width=\columnwidth]{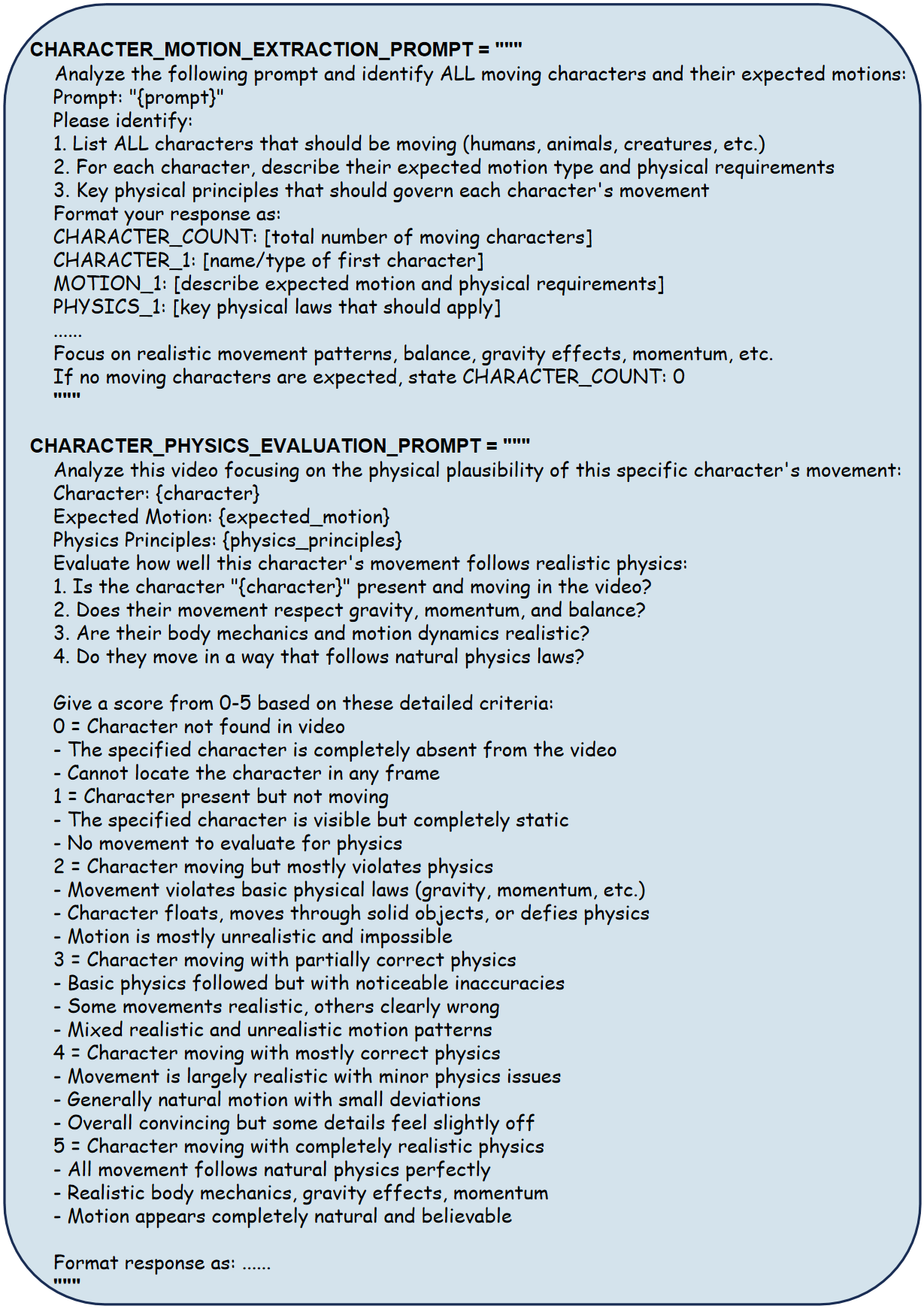}
    \caption{Prompt for Physical Plausibility}
    \label{fig:Physical Plausibility}
\end{figure}

\begin{figure}
    \centering
    \includegraphics[width=\columnwidth]{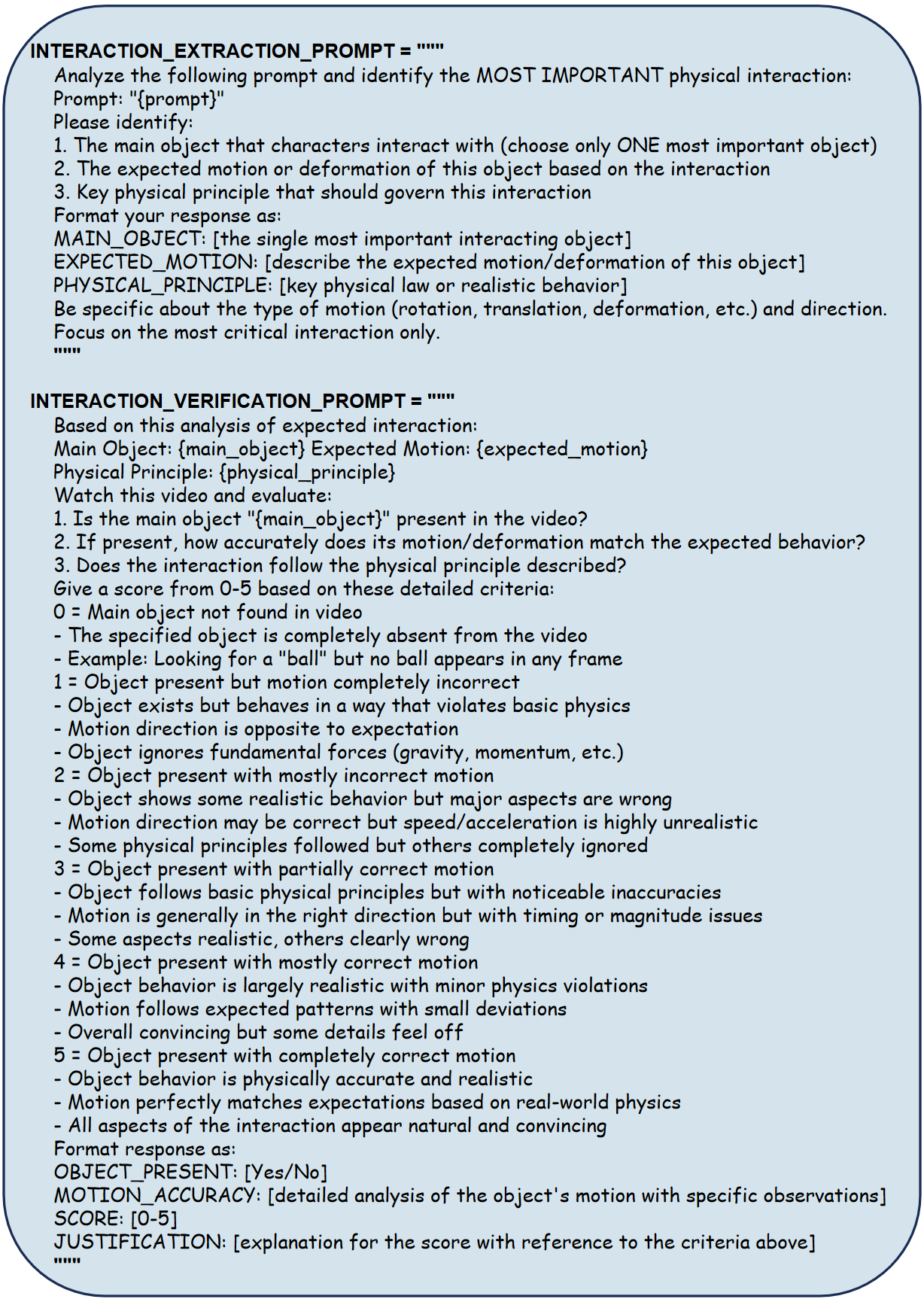}
    \caption{Prompt for Physical Interaction Accuracy}
    \label{fig:Physical Interaction Accuracy}
\end{figure}

To assess Story Video Alignment, Figures \ref{fig:Detect & Count Score} and \ref{fig:Shot Perspective Alignment} illustrate the prompts for verifying whether the generated visual elements and cinematic perspectives align with the textual narrative. The evaluation of Video Consistency is detailed in Figures \ref{fig:State Shift & Persistence} through \ref{fig:Relative Size Consistency}, which focus on character identity, appearance stability, and relative spatial proportions across multiple scenes. Finally, Figures \ref{fig:Action Recognition} through \ref{fig:Physical Interaction Accuracy} provide the prompts tailored for Motion Quality, encompassing action fidelity, camera control, and the plausibility of physical interactions.

\clearpage

\section{Human Rating Details}
\label{sec:HumanRatingDetails}

\begin{figure}
    \centering
    \includegraphics[width=\columnwidth]{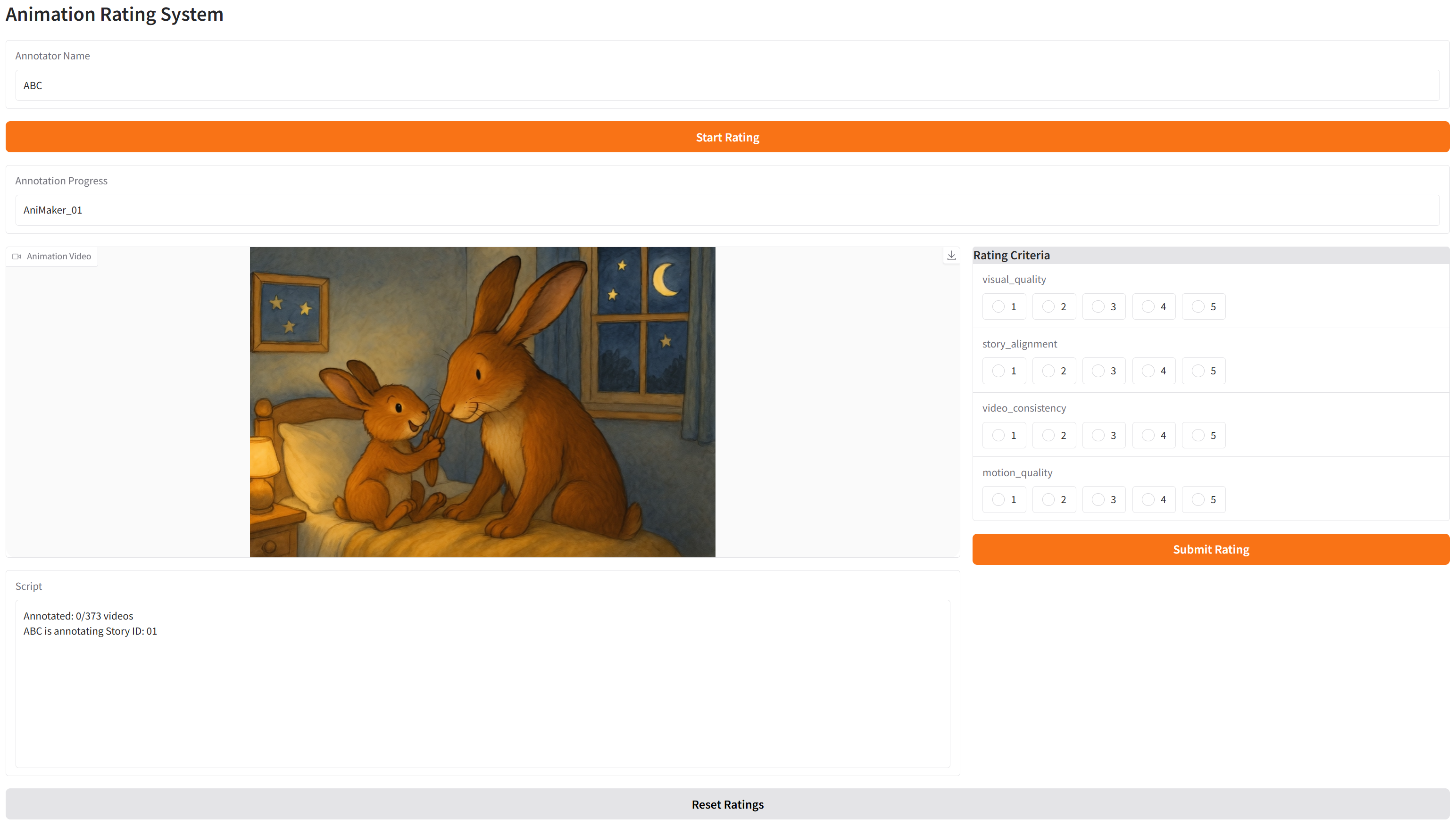}
    \caption{Human Rating System Interface.}
    \label{fig:Human Rating System}
\end{figure}

To evaluate the perceptual quality of the generated videos and validate the alignment between MSVBench and human judgments, we conduct a comprehensive human evaluation involving 10 participants with animation backgrounds. We assess storytelling videos synthesized by 20 models based on MSVBench scripts. As illustrated in Figure~\ref{fig:Human Rating System}, we employ a specialized annotation system and establish rigorous rating guidelines, incorporating pre-evaluation training to ensure inter-rater reliability. Evaluators rate each video on a 5-point scale across four dimensions—Visual Quality, Story Video Alignment, Video Consistency, and Motion Quality—following the specific instructions detailed below.

\subsection{Visual Quality}
Evaluators are instructed to rate the overall aesthetic appeal and technical fidelity of the video content. The focus is on image clarity, lighting distribution, and stylistic unity. Raters must scrutinize the video for technical imperfections, identifying issues such as blurriness, noise, severe distortion, or structural collapse typical of generative artifacts.

\begin{itemize}
    \item \textbf{1 (Unusable):} The video is filled with severe noise, distortion, or artifacts. Lighting is chaotic, and the subject or the scene is totally unrecognizable.
    \item \textbf{2 (Poor):} Visible artifacts (e.g., limb distortion, abnormal color blocks) are present. The image is blurry, or the style fluctuates violently, causing visual discomfort.
    \item \textbf{3 (Fair):} The image is generally clear and recognizable. However, there are localized issues with lighting (e.g., slight overexposure), minor discontinuities between shots, or slight stylistic dissonance between the subject and background.
    \item \textbf{4 (Good):} The video is clear with good aesthetic composition and natural lighting. Minor edge flickering or background blur exists but does not affect the overall viewing experience.
    \item \textbf{5 (Excellent):} The video exhibits cinema-grade illustrative aesthetics with rich lighting layers and a highly unified style. No obvious generation traces are visible.
\end{itemize}

\subsection{Story Video Alignment}
This dimension requires evaluators to determine the semantic correspondence between the generated video and the provided scripts. The assessment involves verifying whether the described characters, objects, and environments are present and possess the correct attributes (e.g., color, quantity). Additionally, evaluators check for adherence to specific camera language (e.g., close-up, overhead view) mentioned in the script.

\begin{itemize}
    \item \textbf{1 (Irrelevant):} The content is unrelated to the prompt, exhibiting severe hallucinations (e.g., generating a city instead of a forest).
    \item \textbf{2 (Severe Mismatch):} The general environment is correct, but the core subject is missing, or fundamental attributes are wrong (e.g., a blue car instead of a red one).
    \item \textbf{3 (Partial Deviation):} The core subject and environment are correct, but there are errors in camera shot types or missing details.
    \item \textbf{4 (High Fidelity):} All major visual elements and attributes are accurately presented. Minor discrepancies are permitted in ambiguous details not explicitly described in the text.
    \item \textbf{5 (Precise Match):} The video is a precise visual translation of the script. All objects, colors, spatial relationships, and specific camera angles are perfectly executed.
\end{itemize}

\subsection{Video Consistency}
Evaluators assess the temporal coherence and stability of visual elements across consecutive frames and shots. The primary task is to detect any unnatural morphing or illogical discrepancies in character identities (appearance, clothing), scene layouts, and relative object scales, to determine the extent to which visual narrative continuity is maintained throughout the sequence.

\begin{itemize}
    \item \textbf{1 (Incoherent):} Character identity or scene layout completely changes after a shot transition, making it impossible to establish narrative continuity.
    \item \textbf{2 (Disjointed):} The character is recognizable between shots, but there are sudden changes in clothing color, severe facial deformation, or jumps in scene style.
    \item \textbf{3 (Average):} Character identity is generally preserved, but details (e.g., patterns, accessories) flicker or change during side profiles, long shots, or large movements.
    \item \textbf{4 (Stable):} Characters and scenes remain stable in most shots. Slight differences are acceptable only under extreme camera angles or dramatic lighting changes.
    \item \textbf{5 (Consistent):} Character features and spatial relationships remain stable throughout the video. There are no perceptible changes in relative scale, maintaining logical consistency regardless of rendering variations.
\end{itemize}

\subsection{Motion Quality}
This criterion evaluates the naturalness, physical plausibility, and semantic accuracy of the depicted dynamics. Evaluators examine whether the actions align with the scripts, if the movements are fluid (free from stiffness or unnatural jitter), and if physical interactions (e.g., gravity, collisions) and camera movements adhere to real-world physical laws.

\begin{itemize}
    \item \textbf{1 (Incorrect / Broken):} Actions are incorrect (e.g., standing still instead of dancing) or violate physics (e.g., floating objects, severe clipping, non-human distortion).
    \item \textbf{2 (Stiff / Unnatural):} The execution deviates significantly from the semantic intent (e.g., mere mouth opening instead of laughter) or exhibits excessive rigidity, resembling a static image translating across the frame.
    \item \textbf{3 (Weak / Incomplete):} The action category is correct, but the magnitude is too small, lacks weight, or is incomplete. Complex interactions often result in clipping.
    \item \textbf{4 (Correct):} Actions largely align with the description; character movement and camera movements are smooth and adhere to ergonomics. Physics are generally correct, with minor flaws allowed only in high-speed or complex interactions.
    \item \textbf{5 (Vivid):} Actions are perfectly aligned with the text and visually impactful. Camera movements are cinematic and stable. Interactions possess real weight and physical feedback.
\end{itemize}

\section{Quantitative Results}
\label{sec:QuantitativeResults}
We present the visualization results for two stories from MSVBench in Figure~\ref{fig:Quantitative1} and Figure~\ref{fig:Quantitative2}. 
To ensure a concise yet representative comparison, we display five sampled shots for each story, showcasing the generation results from all 20 evaluated methods. 
These comparisons highlight performance variations regarding Visual Quality, Story Video Alignment, Video Consistency, and Motion Quality. 
This detailed visualization provides an intuitive understanding of each method's capability in multi-shot video generation.

\begin{figure*}[p] 
    \centering
    \includegraphics[width=0.9\textwidth, height=0.99\textheight, keepaspectratio]{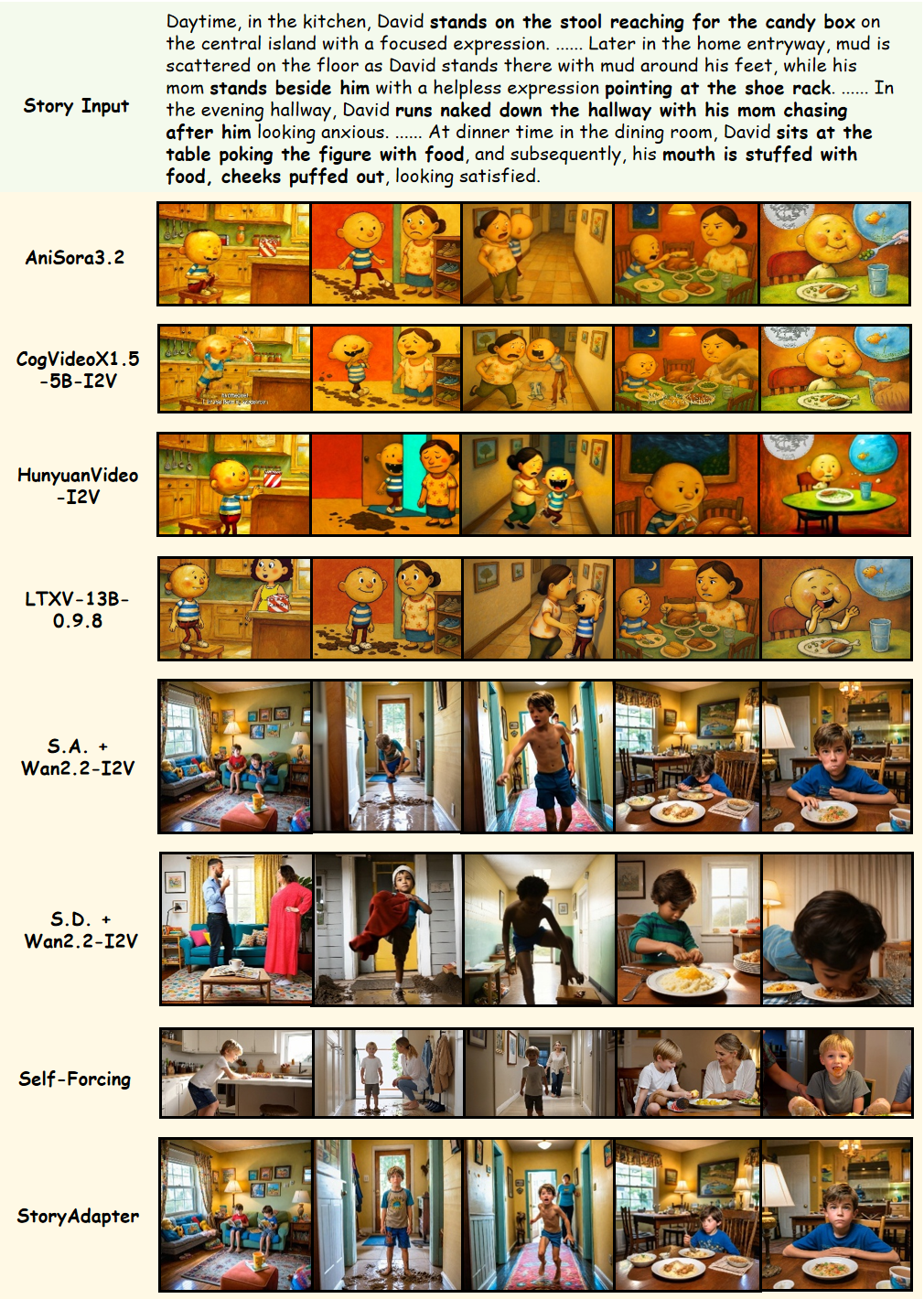}
    
    \caption{Visualization results of Story 2 from \textsc{MSVBench}. (Continued on next page)}
    \label{fig:Quantitative1}
\end{figure*}

\clearpage 

\begin{figure*}[p]
    \ContinuedFloat 
    \centering
    \includegraphics[width=1\textwidth, height=0.99\textheight, keepaspectratio]{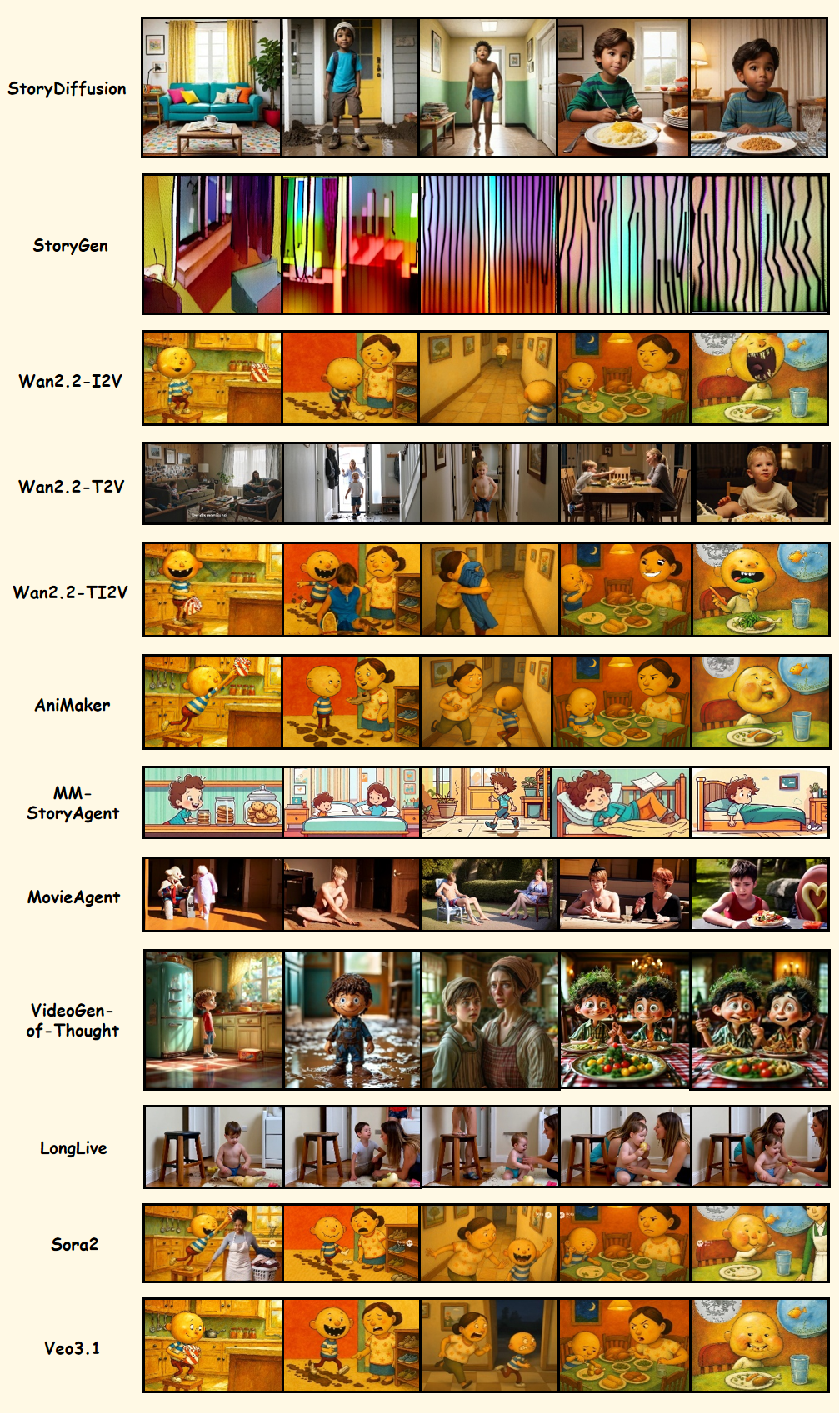}
    
    \caption[]{(Continued)} 
\end{figure*}

\clearpage 

\begin{figure*}[p]
    \centering
    \includegraphics[width=0.9\textwidth, height=0.99\textheight, keepaspectratio]{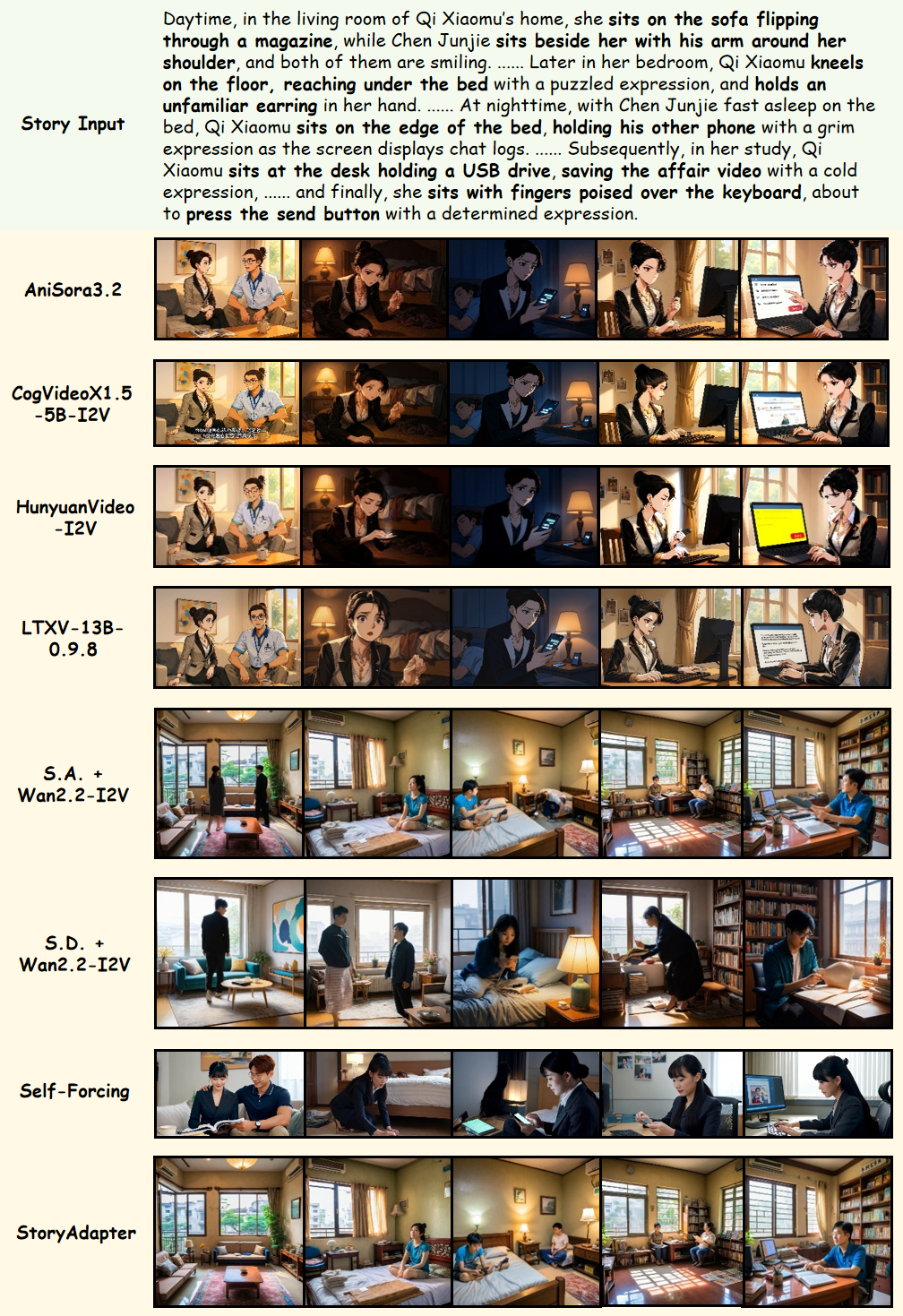}
    
    \caption{Visualization results of Story 14 from \textsc{MSVBench}. (Continued on next page)}
    \label{fig:Quantitative2}
\end{figure*}

\clearpage 

\begin{figure*}[p]
    \ContinuedFloat 
    \centering
    \includegraphics[width=1.0\textwidth, height=0.99\textheight, keepaspectratio]{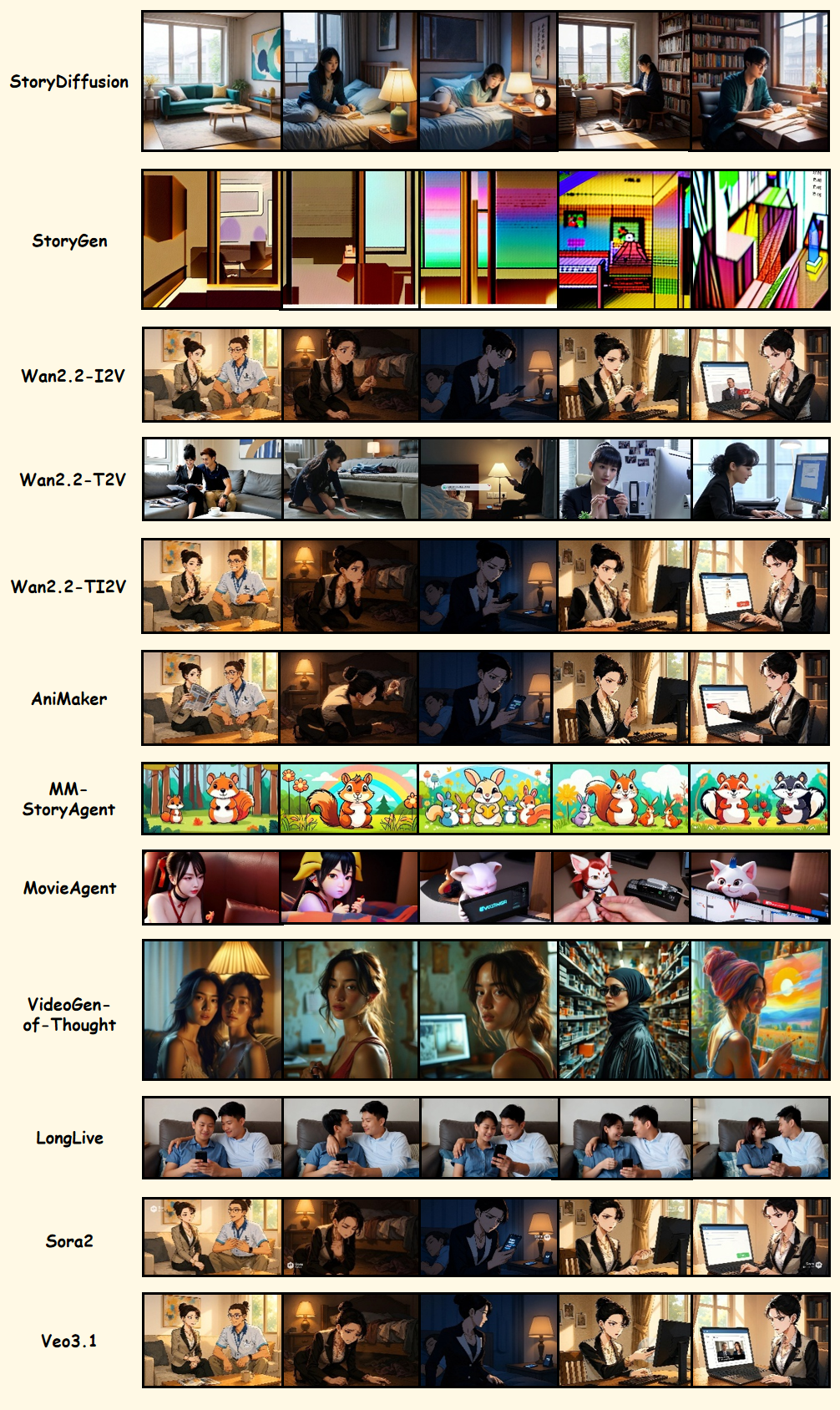}
    
    \caption[]{(Continued)}
\end{figure*}

\begin{figure*}[t!]
  \centering
  \includegraphics[width=\linewidth]{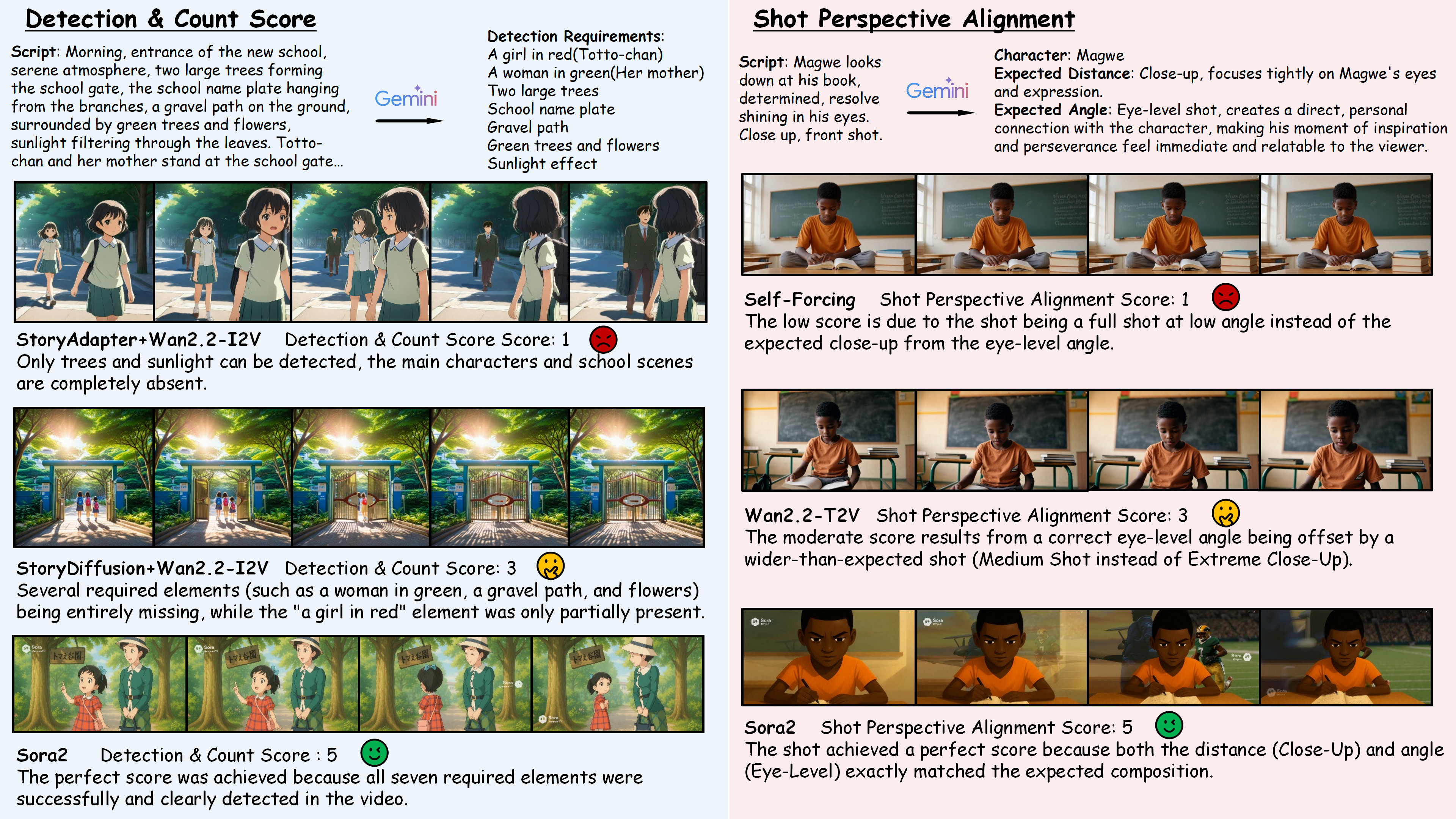} 
  \caption{Case Study on \textbf{Detection \& Count Score} (Left) and \textbf{Shot Perspective Alignment} (Right).}
  \label{fig:case1}
\end{figure*}

\begin{figure*}[t!]
  \centering
  \includegraphics[width=\linewidth]{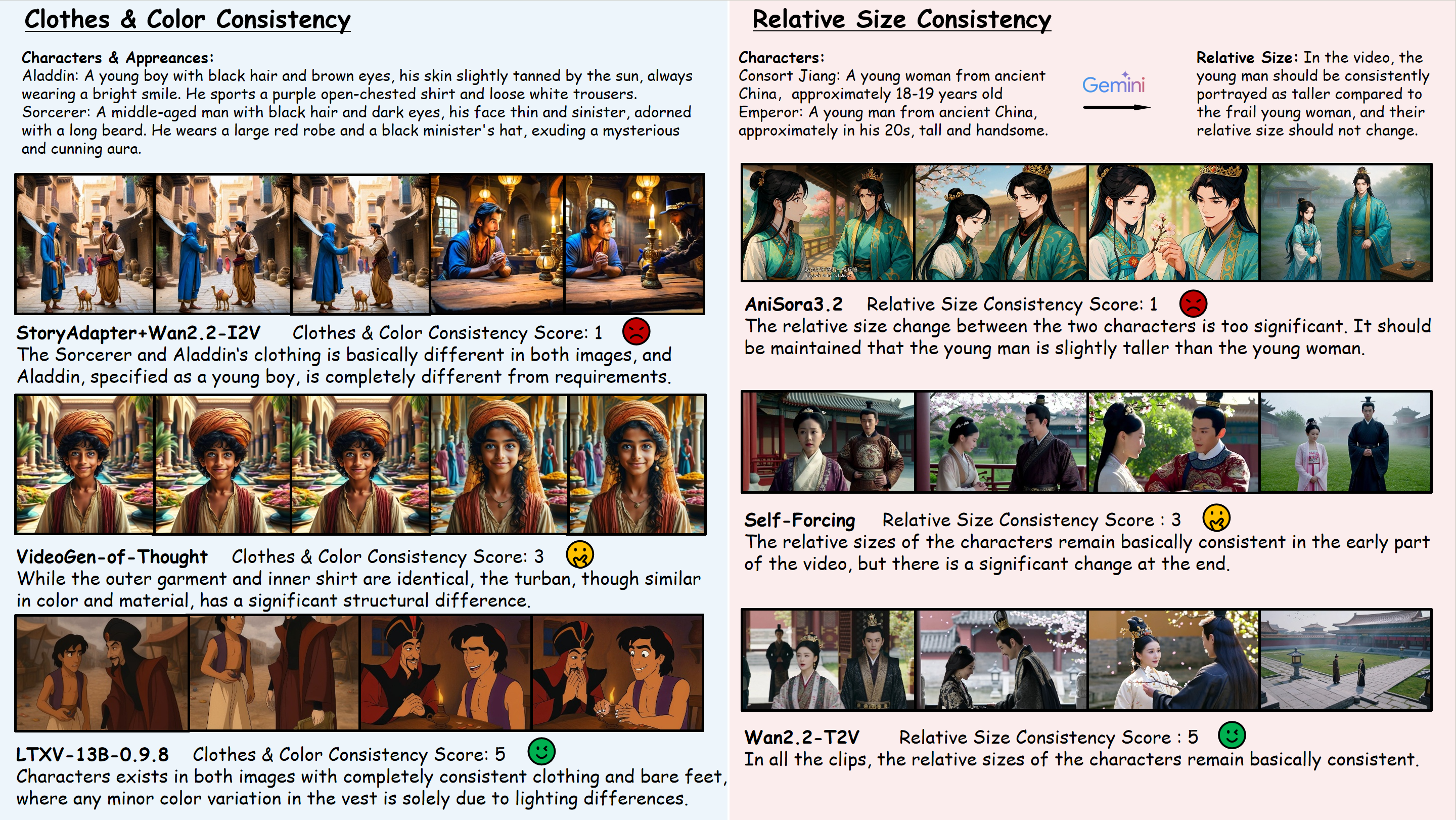} 
  \caption{Case Study on \textbf{Clothes \& Color Consistency} (Left) and \textbf{Relative Size Consistency} (Right).}
  \label{fig:case2}
\end{figure*}

\begin{figure*}[t!]
  \centering
  \includegraphics[width=\linewidth]{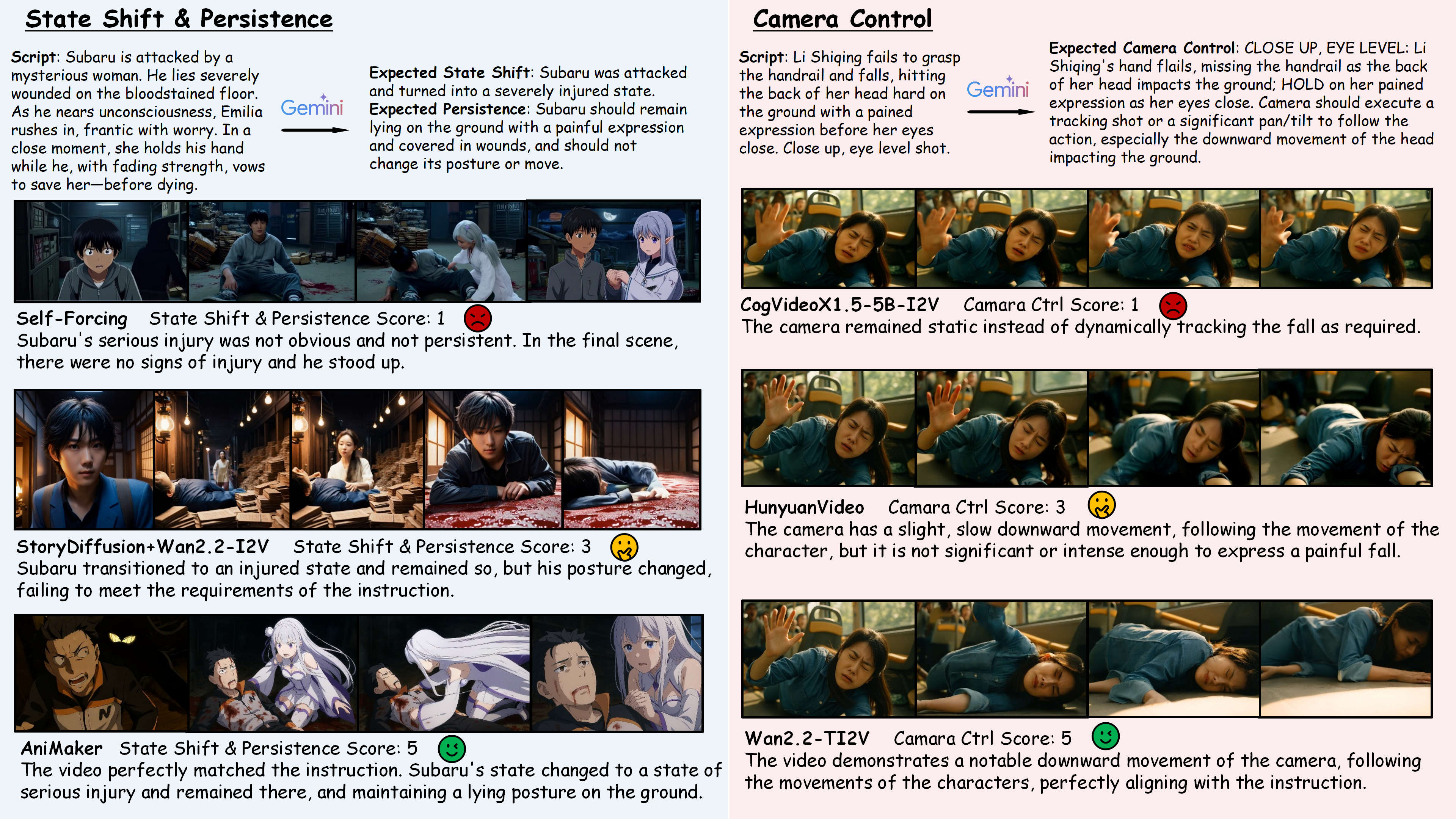} 
  \caption{Case Study on \textbf{State Shift \& Persistence} (Left) and \textbf{Camera Control} (Right).}
  \label{fig:case3}
\end{figure*}

\begin{figure*}[t!]
  \centering
  \includegraphics[width=\linewidth]{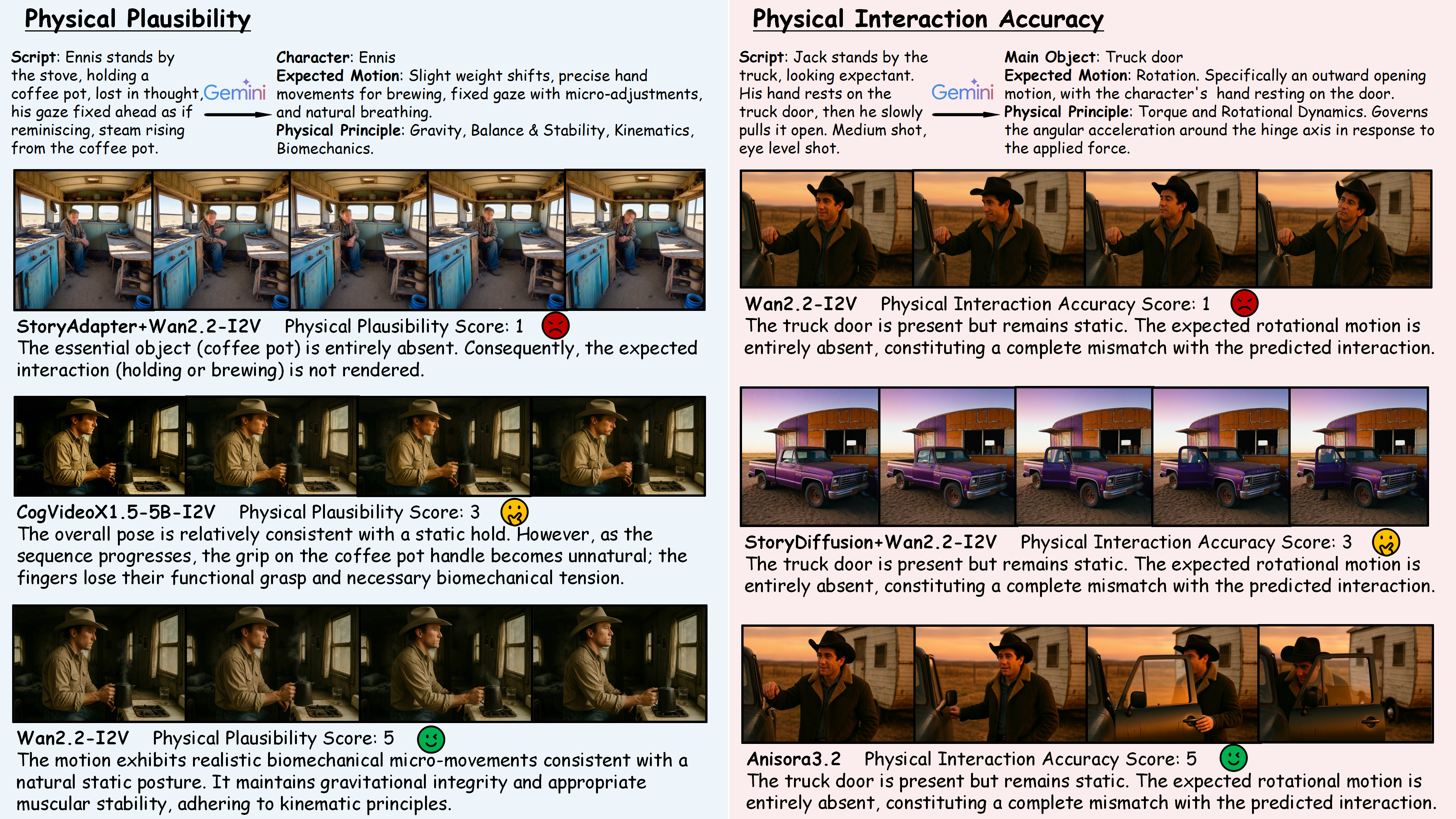} 
  \caption{Case Study on \textbf{Physical Plausibility} (Left) and \textbf{Physical Interaction Accuracy} (Right).}
  \label{fig:case4}
\end{figure*}

\clearpage

\section{Case Study}
\label{sec:CaseStudy}
To explicitly demonstrate the effectiveness and discriminative capability of our proposed MSVBench metrics, we present detailed case studies across all LMM based metrics. Figures~\ref{fig:case1} through \ref{fig:case4} visualize how the LMM extracts semantic requirements from the input scripts and assesses the generated videos.

\subsection{Story Video Alignment Assessment}
Figure~\ref{fig:case1} illustrates the evaluation of \textbf{Detection \& Count Score} and \textbf{Shot Perspective Alignment}.
For detection, the LMM first parses the script to identify necessary entities (e.g., ``A girl in red,'' ``School name plate''). It then verifies both the presence and the precise quantity of these elements in each video; for instance, it assigns a low score to StoryAdapter (Score 1) for failing to generate required characters, while confirming that Sora2 (Score 5) achieves perfect recall of all mandated elements.
For cinematography, the evaluator assesses the realized shot against the intended camera language without reference to other models. As shown, it accurately penalizes Self-Forcing (Score 1) for generating a full shot instead of the required close-up, while validating Sora2's (Score 5) perfect adherence to the ``Close-Up'' and ``Eye-Level'' constraints.

\subsection{Video Consistency Assessment}
Figure~\ref{fig:case2} demonstrates the \textbf{Clothes \& Color Consistency} and \textbf{Relative Size Consistency} metrics.
In multi-shot generation, maintaining visual attributes is critical. The LMM identifies specific violations in individual model outputs, such as the structural change in the turban in VideoGen-of-Thought (Score 3) or complete clothing mismatches in StoryAdapter (Score 1). Conversely, it acknowledges the consistent output generated by LTXV (Score 5). Furthermore, the Relative Size metric detects logical inconsistencies in character proportions across shots, such as penalizing AniSora3.2 (Score 1) for significantly altering the height difference between the protagonist and the consort.

\subsection{Complex Logic and Camera Control}
Figure~\ref{fig:case3} showcases the evaluation of temporal logic via \textbf{State Shift \& Persistence} and \textbf{Camera Control}.
The State Shift metric evaluates whether narrative-driven changes (e.g., an injury) persist logically within a specific video. The LMM successfully identifies that Self-Forcing (Score 1) violates the narrative when the character "Subaru" stands up uninjured in the final scene, whereas it recognizes that AniMaker (Score 5) successfully maintains the "severely injured" state throughout.
Simultaneously, the Camera Control metric assesses dynamic movement against the prompt. The evaluator assigns a low score to CogVideoX (Score 1) for producing a static shot, while separately validating the complex tracking shot in Wan2.2-TI2V (Score 5), demonstrating its sensitivity to motion dynamics.

\subsection{Motion Quality and Physics}
Figure~\ref{fig:case4} focuses on \textbf{Physical Plausibility} and \textbf{Physical Interaction Accuracy}.
These metrics assess adherence to real-world physics for each generation. For plausibility, the LMM detects subtle biomechanical issues, such as the unnatural grip on a coffee pot in CogVideoX (Score 3), and separately confirms the natural static posture in Wan2.2 (Score 5). Regarding interactions, the system verifies causal logic; for instance, it penalizes Wan2.2-I2V (Score 1) where a ``truck door'' remains static despite a prompt explicitly describing it opening, while recognizing the correct rotational dynamics in other successful generations like Anisora3.2 (Score 5).

Collectively, these cases above confirm that MSVBench provides fine-grained, objective scores for each model. By evaluating generated content strictly against the input script and physical principles rather than performing relative rankings, the framework effectively penalizes hallucinations, continuity errors, and physical violations.

\section{Correlation Coefficient Details}
\label{sec:CorrelationCoefficientDetails}
To evaluate the alignment between the proposed MSVBench metrics and human perception, we compute correlation coefficients between objective metric scores and human ratings. This involves three steps: (1) metric aggregation, which normalizes raw sub-metrics into ranks to compute dimension-level scores via negative averaging; (2) statistical analysis, utilizing Spearman’s rank correlation and Kendall’s rank correlation coefficients to quantify the alignment between MSVBench and human ratings; and (3) overall score calculation, which derives a unified model ranking by averaging the scores across the four evaluation dimensions.

\subsection{Metric Aggregation}
Given that MSVBench sub-metrics vary widely in scale and units, we normalize them by ranking.  
For $N$ models, let $v_{i,k}$ denote the raw value of the $k$-th sub-metric for model $i$.  
Each value is converted into a rank $r_{i,k} \in [1, N]$, where $r=1$ represents the best performance.  
If multiple models share the same value, we assign them the average rank of the tied group.

For a dimension (e.g., Visual Quality) with $K$ sub-metrics, the aggregated dimension-level score for model $i$ is:
\begin{equation}
    M_i = - \frac{1}{K} \sum_{k=1}^{K} r_{i,k},
\end{equation}
where the negative sign ensures that higher-quality models (with smaller ranks) receive larger $M_i$ values, aligning the rating direction with human evaluation scores.

\subsection{Statistical Analysis}
We employ two non-parametric rank-based correlation coefficients to quantify the alignment between MSVBench scores and human ratings. Spearman's rank correlation ($\rho$) is utilized to assess the global monotonic consistency between the two ranking lists, reflecting how well the overall ranking trend matches human perception. Complementarily, Kendall's rank correlation ($\tau$) focuses on the accuracy of pairwise relative orderings, providing a more rigorous evaluation of measuring ordinal association and robustness against ranking noise. 

\noindent \textbf{Spearman’s Rank Correlation ($\rho$):}  
Let $\hat{r}_{m,i} = R(M_i) - \bar{R}_M$ and $\hat{r}_{h,i} = R(H_i) - \bar{R}_H$ denote centered ranks for MSVBench and human scores, respectively. The coefficient is:
\begin{equation}
    \rho = 
    \frac{
        \sum_i 
        \hat{r}_{m,i}\,
        \hat{r}_{h,i}
    }{
        \sqrt{
            \sum_i \hat{r}_{m,i}^2
        }
        \sqrt{
            \sum_i \hat{r}_{h,i}^2
        }
    }.
\end{equation}
A high $\rho$ indicates consistent ordering between MSVBench and human judgments.

\noindent \textbf{Kendall's Rank Correlation ($\tau$):}  
Using pairwise consistency,
\begin{equation}
    \tau = 
    \frac{
        N_c - N_d
    }{
        \sqrt{
            (N_0 - N_1)(N_0 - N_2)
        }
    },
\end{equation}
where $N_c$ and $N_d$ are concordant and discordant pairs,  
$N_0 = \frac{N(N-1)}{2}$ is the total number of model pairs,  
and $N_1$, $N_2$ correct for ties in MSVBench and human scores, respectively.  
This formulation ensures robust ranking comparison when ties occur.

\subsection{Overall Score Calculation}
Besides dimension-level evaluation, MSVBench provides a unified \textit{overall score} for each model.  
For each model $i$, let $M^{\text{VQ}}_i$, $M^{\text{SA}}_i$, $M^{\text{VC}}_i$, and $M^{\text{MQ}}_i$ denote the aggregated (negative) rank scores of the four MSVBench dimensions.  
The overall score, also referred to as the model's \textbf{Average Rank}, is defined as the mean of these four dimension-level rank scores:
\begin{equation}
    R^{\text{overall}}_i = -\frac{1}{4}
    \left(
        M^{\text{VQ}}_i +
        M^{\text{SA}}_i +
        M^{\text{VC}}_i +
        M^{\text{MQ}}_i
    \right).
\end{equation}

\begin{table}[t]
\centering
\caption{MSVBench Leaderboard: Overall average ranking of all evaluated methods.}
\label{tab:overall_ranking}
\begin{tabular}{rlc}
\toprule
Rank & Model & Overall Ranking$^\downarrow$ \\
\midrule
 1 & Veo3.1                   & $5.60$ \\
 2 & Sora2                    & $6.67$ \\
 3 & Wan2.2\text{-}TI2V       & $7.50$ \\
 4 & Anisora3.2               & $7.90$ \\
 5 & AniMaker                 & $8.00$ \\
 6 & Wan2.2\text{-}T2V        & $8.12$ \\
 7 & Wan2.2\text{-}I2V        & $8.45$ \\
 8 & Self-Forcing             & $8.57$ \\
 9 & LTXV-13B-0.9.8           & $9.15$ \\
10 & HunyuanVideo\text{-}I2V  & $9.60$ \\
11 & CogVideoX1.5\text{-}5B\text{-}I2V & $9.87$ \\
12 & LongLive                 & $10.21$ \\
13 & S.D.+Wan2.2\text{-}I2V   & $10.95$ \\
14 & S.A.+Wan2.2\text{-}I2V   & $11.05$ \\
15 & VideoGen-of-Thought      & $11.42$ \\
16 & StoryDiffusion           & $12.17$ \\
17 & StoryAdapter             & $12.42$ \\
18 & MovieAgent               & $15.19$ \\
19 & MM-StoryAgent            & $15.30$ \\
20 & StoryGen                 & $16.70$ \\
\bottomrule
\end{tabular}
\end{table}

Table~\ref{tab:overall_ranking} presents the MSVBench Leaderboard. Notably, the derived overall ranking demonstrates strong alignment with human perception, validating the reliability of our MSVBench.

\end{document}